\documentclass[%
 preprint,
 amsmath,
 amssymb,
 aps, 
]{revtex4-2}

\usepackage{graphicx}
\usepackage{dcolumn}
\usepackage{bm}

\usepackage{xcolor}

\usepackage{physics}
\usepackage{capt-of}
\usepackage{tabularx} 
\usepackage{xfrac}
\newcommand\Tstrut{\rule{0pt}{2.6ex}}       
\newcommand\Bstrut{\rule[-1.2ex]{0pt}{0pt}} 

\begin{document}


\title{Near-wall depletion and layering affect contact line friction of multicomponent liquids}

\author{Michele Pellegrino}

\author{Berk Hess}%
\affiliation{Swedish e-Science Research center, Science for Life Laboratory, Department of Applied Physics KTH, Stockholm 100 44, Sweden}



\date{\today}

\begin{abstract}
    The main causes of energy dissipation in micro- and nano-scale wetting are viscosity and liquid-solid friction localized in the three-phase contact line region. Theoretical models predict the contact line friction coefficient to correlate with the shear viscosity of the wetting fluid. Experiments conducted to investigate such correlation have not singled out a unique scaling law between the two coefficients. We perform Molecular Dynamics simulations of liquid water-glycerol droplets wetting silica-like surfaces, aimed to demystify the effect of viscosity on contact line friction. The viscosity of the fluid is tuned by changing the relative mass fraction of glycerol in the mixture and it is estimated both via equilibrium and non-equilibrium Molecular Dynamics simulations. Contact line friction is measured directly by inspecting the velocity of the moving contact line and the microscopic contact angle. It is found that the scaling between contact line friction and viscosity is sub-linear, contrary to the prediction of Molecular Kinetic Theory. The disagreement is explained by accounting for the depletion of glycerol in the near-wall region. A correction is proposed, based on multicomponent Molecular Kinetic Theory and the definition of a re-scaled interfacial friction coefficient.
\end{abstract}

\keywords{Molecular Dynamics, Wetting, Droplets, Moving contact lines, Nanofluidics, Viscosity} 
\maketitle

\section{Introduction}
\label{sec:intro}

The dynamics of wetting and dewetting involves the motion of three-phase contact lines, which is thus ubiquitous both in nature and engineering. Deepening the physical understanding of moving contact lines is therefore essential in order to improve industrial processes such as coating \cite{weinstein2004coating}, 3D-printing \cite{deruiter2017print3d} and boiling \cite{bures2022boiling}, as well as to provide insight for the development of bio-inspired surfaces \cite{solga2007bioinspired,mouterde2017cicada}. The research on this topic involves expertise from several communities and has produced a plethora of mathematical models \cite{voinov1976model, cox1986viscous, petrov1992combined, shikhmurzaev1997ift,blake2006physics}. Despite differences between modeling approaches, it is clear that viscosity and liquid-solid friction represent the main sources of energy dissipation, and thus dominate the dynamics of contact lines at sufficiently small length scales \cite{doquang2015wetting}. 

The effect of liquid-solid friction localized in the three-phases region, i.e. contact line friction, has been studied extensively and it has been determined to produce a deviation of the dynamic contact angle from its Young-Dupr\'{e} equilibrium value \cite{carlson2012universal,johansson2015physicochemistry,johansson2019friction}. The prevalent first-principle explanation of contact line friction emerges from the Molecular Kinetic Theory (MKT) formulated by Blake and Haynes \cite{blake1969mkt}: when a liquid-vapor interface impinges a flat solid surface, the energy balance in the displacement of a three-phase contact line involves on one hand the work of adhesion of liquid molecules on the solid surface and on the other hand the attraction forces between liquid molecules in the near-wall region. The latter component is in turn postulated to correlate with the viscosity of the wetting liquid. It entails that the contact line friction coefficient for a fluid-solid combination scales linearly with the shear viscosity coefficient \cite{blake2002interaction}.

Recently, numerous experimental studies have probed the applicability and the implications of MKT \cite{blake2002shikhmurzaev,ramiasa2013defects,andrukh2014meniscus,karim2016forcedspontaneous,barriozhang2020friction}. For instance, Duvivier et al. observed the predicted linear scaling between viscosity and contact line friction in the spreading dynamics of water-glycerol droplets over glass \cite{duvivier2011experiment}. The scaling was confirmed by a following work which analyzed 20 separate dynamic contact angle studies \cite{duvivier2013predictive}. The experiments on silica involved estimating the contact line friction coefficient directly from optical measurements of the dynamic contact angle \cite{seveno2009wetting}; this approach might not fully capture the bending effect of friction on the liquid-vapor interfaces, which extends below the optical resolution limit \cite{chen2014nanobending,deng2016experimental}. Carlson et al. pursued a different approach and inferred the value of contact line friction by reproducing the spreading rate of water-glycerol droplets on silica by simulating Cahn–Hilliard-Navier-Stokes equations with ad hoc wetting boundary conditions \cite{carlson2012universal,carlson2012contactline}. In disagreement with MKT, the scaling between contact line friction and viscosity that matched simulation results was found to be sub-linear. More recently, Li et al. estimated the liquid-solid friction coefficient indirectly from spreading rates and found no linear correlation with viscosity \cite{li2023friction}. The disagreement between experimental studies indicates the limitations of optical microscopy in disentangling the effect of viscous bending at the visible scale (\textit{apparent} contact angle) from the effect of contact line friction at the molecular scale (\textit{microscopic} contact angle).

Recent advances in molecular dynamics simulations have provided methods to directly observe and analyze the nanoscopic dynamics of moving contact lines via numerical experiments \cite{johansson2019friction,papadopoulou2019graphene,toledano2020hidden,ozsipahi2022nanochannel,malgaretti2022hydrodynamics}. Nevertheless, no molecular dynamics study has been focused on studying the correlation between viscosity and contact line friction. In this paper we intend to investigate the aforementioned mismatch between experimental evidence by employing molecular simulations to carefully estimate viscosity and contact line friction of water-glycerol droplets wetting hydrophilic surfaces. Rheological studies have found that the shear viscosity coefficient of water-glycerol mixtures changes drastically depending on the mass fraction of glycerol, while equilibrium interfacial and wetting properties are affected to a lesser extent \cite{takamura2012glycerol,yada2023prf}. Therefore simulating water-glycerol mixtures allows us to isolate and control the effect of viscosity, in addition to imitating a fluid-surface combination utilized in real-world experiments.

The material properties of simulated water-glycerol solutions are found to be generally consistent with experimental results and theoretical prediction. This holds particularly true for the shear viscosity coefficient. Moreover, upon simulating two-dimensional droplet spreading we observe a clear 1:1 relation between dynamic contact angle and contact line speed, which is a signature of contact line friction. Albeit the contact line friction coefficient is higher the higher the viscosity, the scaling law is sub-linear. Upon inspecting the liquid density structure in the near-wall region it is found that glycerol depletes the solid surface, allowing water to form an adsorbed layer. This simple observation explains the sub-linear scaling qualitatively. A quantitative explanation is obtained by considering multicomponent MKT \cite{liang2010multicomponent}; by simply correcting for the local near-wall density of each fluid component, the linear scaling is recovered.


The paper is organized as follows. In section \ref{sec:clfriction} we briefly introduce the linear contact line mobility model based on contact line friction. In section \ref{sec:mdsim} we illustrate the types of Molecular Dynamics simulations performed in this work and their goals. In section \ref{sec:analysis} we present the results of molecular simulations, in particular the calculation on viscosity and contact line friction. In section \ref{sec:discussion} we discuss the implication of simulation results, focusing on the molecular effects at the liquid-solid interface and the comparison with experimental studies. Finally, we draw our conclusions in section \ref{sec:conclusions}.

\section{Contact line friction model}
\label{sec:clfriction}

\begin{figure}[htbp]
    \centering
    \includegraphics[width=0.95\textwidth,trim={0 0 0 0},clip]{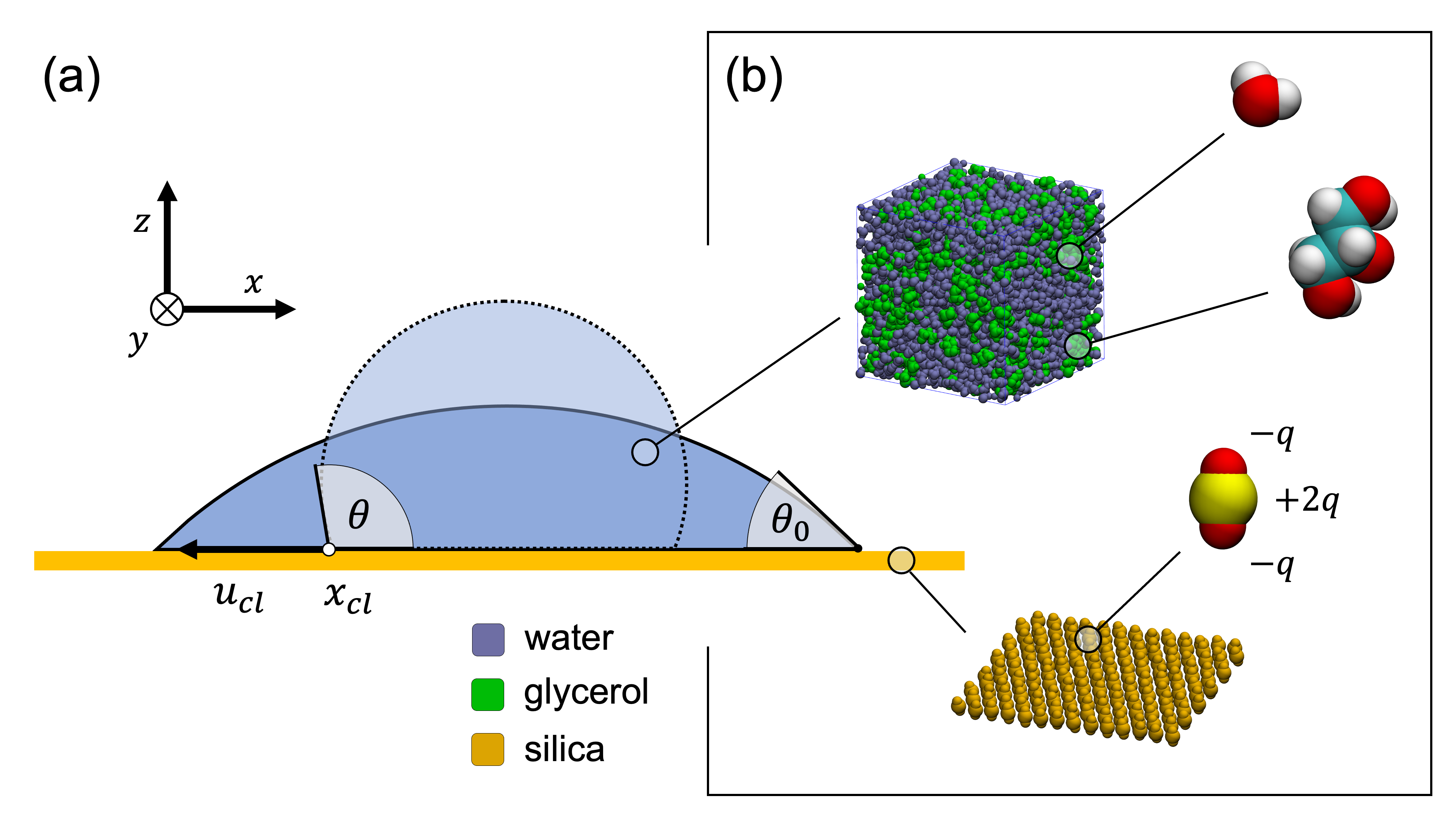}
    \caption{(a) Schematic representation of the moving contact line of a two-dimensional spreading droplet. The reported reference system is the one adopted in molecular simulations. The opaque droplet represents the initial configuration and the transparent one the final equilibrium configuration. (b) Molecular details of the liquid-solid combination, reporting the location of the partial charge $q$ of silica quadrupoles.}
    \label{fig:molecular_scheme}
\end{figure}

Contact line motion is modeled in a two-dimensional geometry. We also assume that one of the two fluids is dense and the other is its vapor phase. The contact point where the liquid-vapor impinges on the solid surface moves with velocity $u_{cl}$ in the direction parallel to the solid surface (figure \ref{fig:molecular_scheme}a). The contact angle $\theta$ is allowed to deviate from its equilibrium value $\theta_0$. Molecular Kinetic Theory expresses the velocity of the contact point in terms of frequency and length of discrete molecular displacement events. For small differences between the equilibrium and the dynamic contact angle, or equivalently for sufficiently small contact line velocities, the following linear mobility relation is usually adopted:

\begin{equation}    \label{eq:linear-mkt}
    \mu_f u_{cl} = \Big(\frac{k_BT}{\kappa^0\lambda^3}\Big) u_{cl} = \sigma\big(\cos\theta_0-\cos\theta\big) \; ,
\end{equation}

$\mu_f$ being referred to as the \textit{contact line friction} coefficient and $\sigma$ being the liquid-vapor interface tension; $\lambda$ is the length of molecular jumps and $\kappa^0$ is the equilibrium molecular jump rate. 

The equilibrium jump rate can be reasonably assumed to depend on a per-molecule jump activation free energy $\Delta g$ according to Arrhenius equation: $k^0\propto\exp\{-\Delta g/(k_BT)\}$. The jump activation free energy can be split into one component that quantifies the adsorption (desorption) work on (respectively from) the solid surface $\Delta g_s$, and one encompassing the interactions between neighboring liquid molecules $\Delta g_l$ \cite{blake2002interaction}. The latter is in turns related to the viscosity of the wetting fluid $\eta$ according to Eyring theory: $\eta\propto\exp\{\Delta g_l/(k_BT)\}/v$, being $v$ the volume of a fluid molecule \cite{collins1957visco}. Hence, the equilibrium jump rate is decomposed as:

\begin{equation}    \label{eq:rate-fun-visco}
    \kappa^0 = \kappa_s^0\kappa_l^0 \propto \frac{\kappa_s^0}{\eta v} \; ,
\end{equation}

Note that $\kappa_s^0$ and $\kappa_l^0$ should not be interpreted as jump rates themselves, but simply as characteristic frequencies determined by the work of adsorption on the substrate on one hand and on the other hand by viscosity. Combining equations \ref{eq:linear-mkt} and \ref{eq:rate-fun-visco}, the contact line friction coefficient is found to scale linearly with the shear viscosity coefficient:

\begin{equation}
    \mu_f \propto \frac{k_B T}{\kappa_s^0}\frac{v}{\lambda^3} \eta \; .
\end{equation}

The simplest way to obtain the contact line friction coefficient from molecular simulations of dynamic wetting is to fit equation \ref{eq:linear-mkt} to the contact line speed and the dynamic contact angle, which are both extracted from the location and the shape of the liquid-vapor interface. In order to avoid over-fitting, the equilibrium contact angle and the surface tension can be obtained independently from simple equilibrium simulations. The shear viscosity coefficient can also be obtained independently, although it necessitates particular care, as will be illustrated in the following section.

\section{Methods}
\label{sec:mdsim}

\subsection{Molecular Dynamics simulations}

In this section we briefly describe the molecular simulations performed to measure the material parameters of aqueous glycerol droplets and to study how they wet silica-like substrates. All simulations are run with GROMACS 2022 \cite{gromacs2015}. Additional information on the force field and simulation parameters are provided in \ref{sec:md}. 

The SPC/E model is used to parameterize water molecules, while the force field parameters of glycerol molecules are taken from OPLS-AA according to the study by Jahn et al. \cite{jorgensen1996oplsaa,jahn2014glycerol}. The solid substrate consists of a monolayer of silica quadrupoles arranged in a hexagonal lattice with spacing 0.45 nm (figure \ref{fig:molecular_scheme}b). Oxygen atoms in each quadrupole molecule are partially charged with charge $-q$, to which corresponds a charge $2q$ on silicon atoms. The lattice monolayer geometry does not reproduce the topology of real silica, which can either be a three-dimensional crystal or amorphous, but emulates the electrostatic interactions that are fundamental to correctly reproduce wetting of polar liquids such as water or glycerol. The partial charge $q$ can be tuned to change the wettability of the surface, and thus obtain different contact angles.

Liquid mixtures are produced at six different mass fractions of glycerol, indicated with $\alpha_g=\{0.0, 0.2, 0.4, 0.6, 0.8, 1.0\}$. Cubic liquid boxes are obtained by randomly inserting a fixed number of water and glycerol molecules, according to the mass fraction. Boxes are then equilibrated at constant hydrostatic pressure and temperature first (NPT), and then at constant volume and temperature (NVT), with $T=300$ K and $P=1$ bar. Production runs are performed at constant volume.

\begin{table}
    \centering
    \begin{tabularx}{\textwidth}{| X | l >{\raggedright}p{0.35\textwidth} c l |}
        \hline
        Scheme & Label & Goal & & $L^0_x\times L^0_y\times L^0_z$ \Tstrut\\
        & & & & [nm$^3$] \Bstrut\\
        \hline
        \hline
        \noindent\parbox[c]{\hsize}{\includegraphics[scale=0.09,trim={0 1.25cm 0 1.25cm},clip]{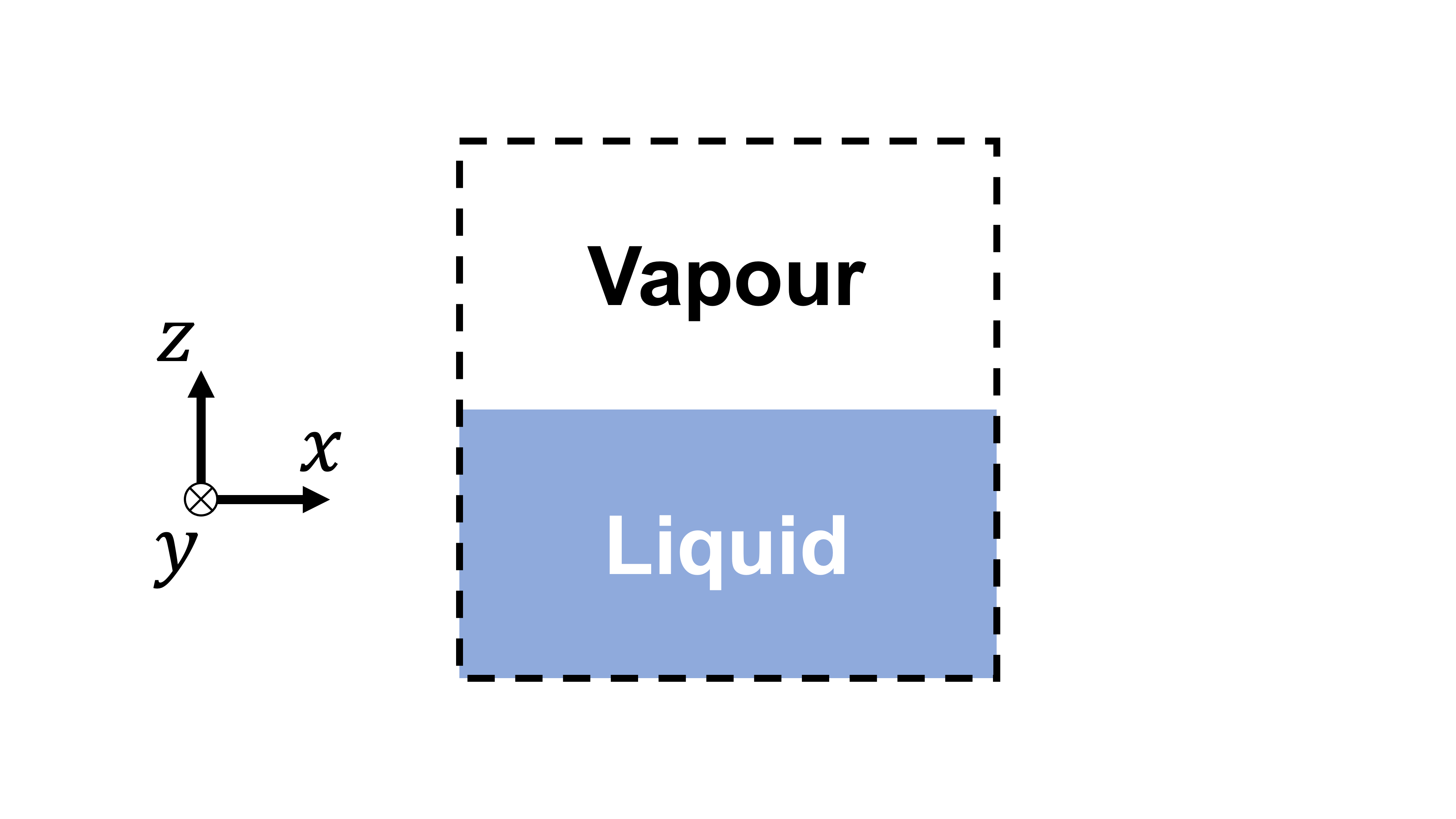}} & SLAB$\quad$ & Surface tension &  & $10\times5\times10$ \\
        \noindent\parbox[c]{\hsize}{\includegraphics[scale=0.09,trim={0 0.75cm 0 0.75cm},clip]{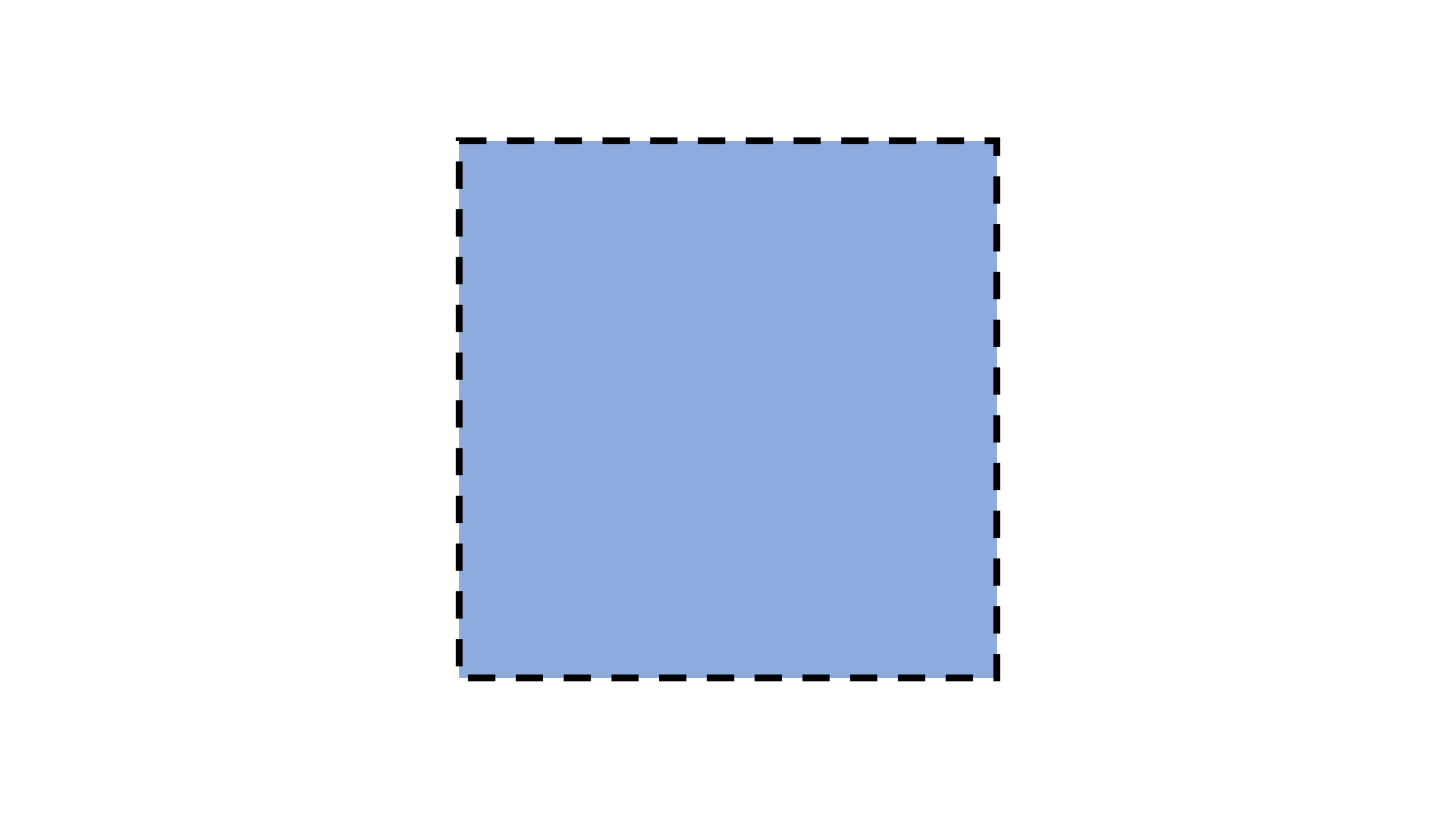}} & BOX I$\quad$ & Viscosity (equilibrium) & & $5\times5\times5$ \\
        \noindent\parbox[c]{\hsize}{\includegraphics[scale=0.09,trim={0 0 0 0},clip]{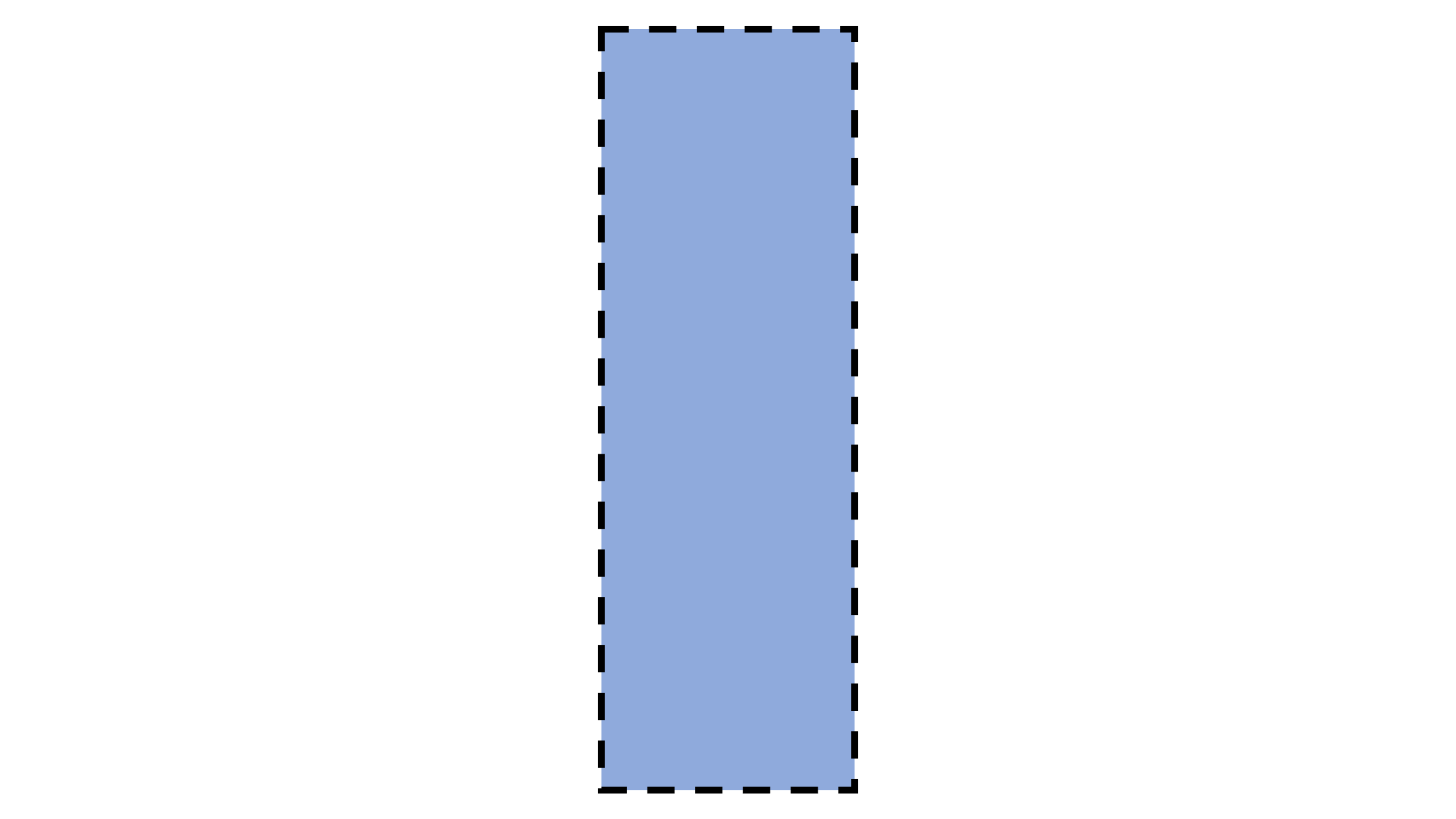}} & BOX II$\quad$ & Viscosity (non-equilibrium) & & $5\times5\times15$ \\
        \noindent\parbox[c]{\hsize}{\includegraphics[scale=0.09,trim={0 0.5cm 0 0.5cm},clip]{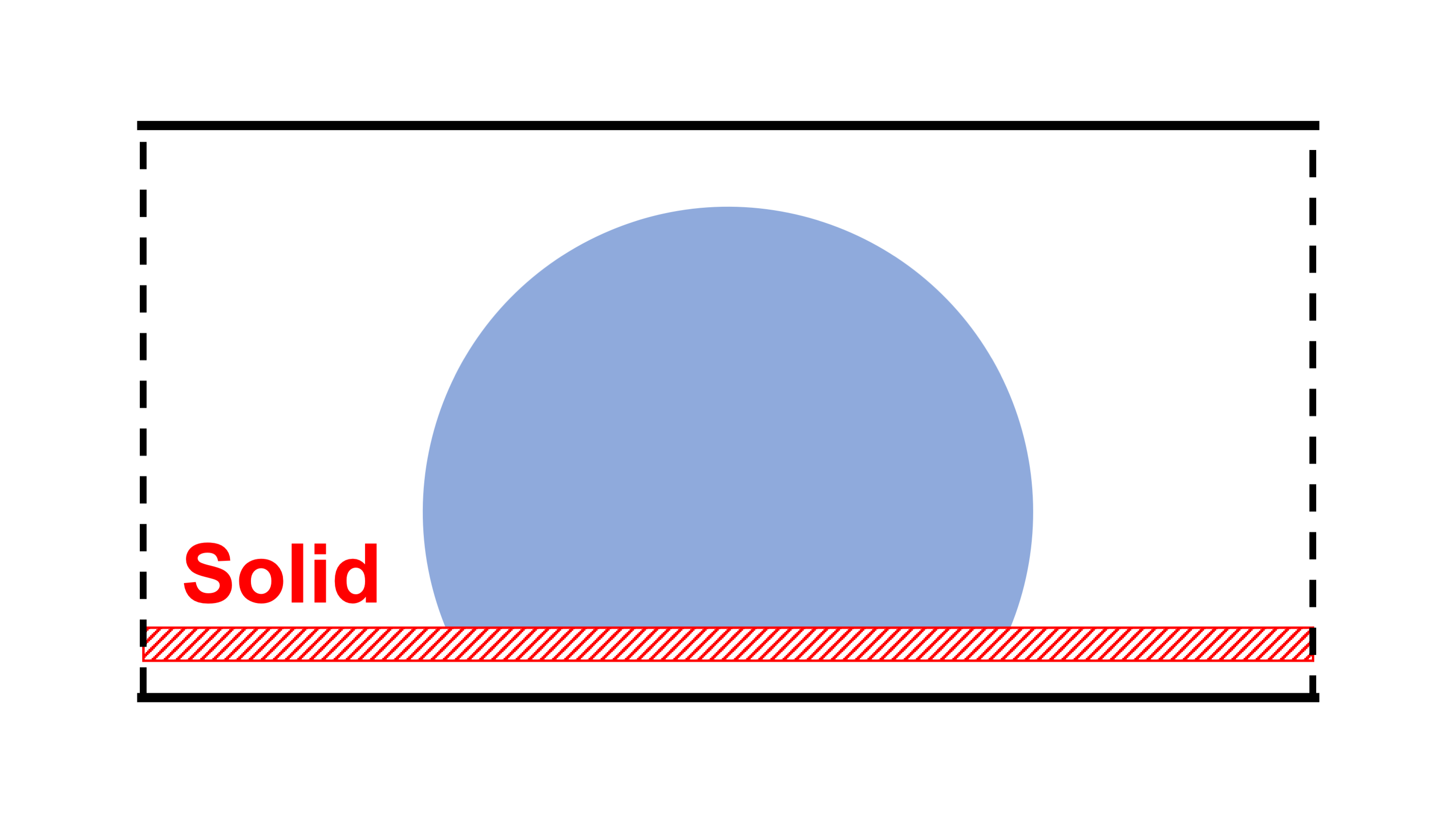}} & DROP I$\quad$ & Equilibrium contact angle & & $60\times4.7\times30$ \\
        \noindent\parbox[c]{\hsize}{\includegraphics[scale=0.09,trim={0 0.5cm 0 0.5cm},clip]{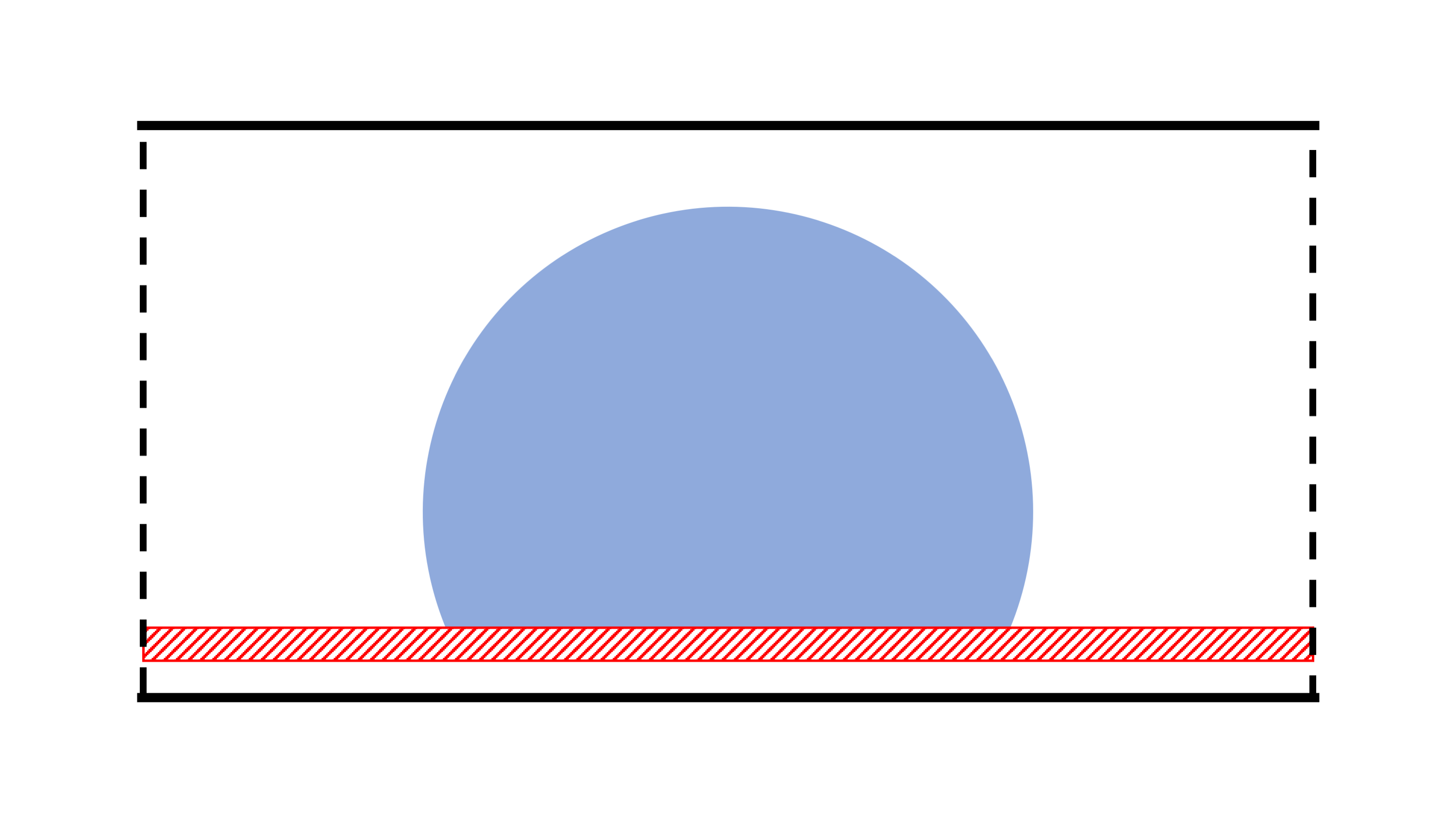}} & DROP II$\quad$ & Contact line friction & & $120\times4.7\times45$ \\
        \noindent\parbox[c]{\hsize}{\includegraphics[scale=0.09,trim={0 0.5cm 0 0.5cm},clip]{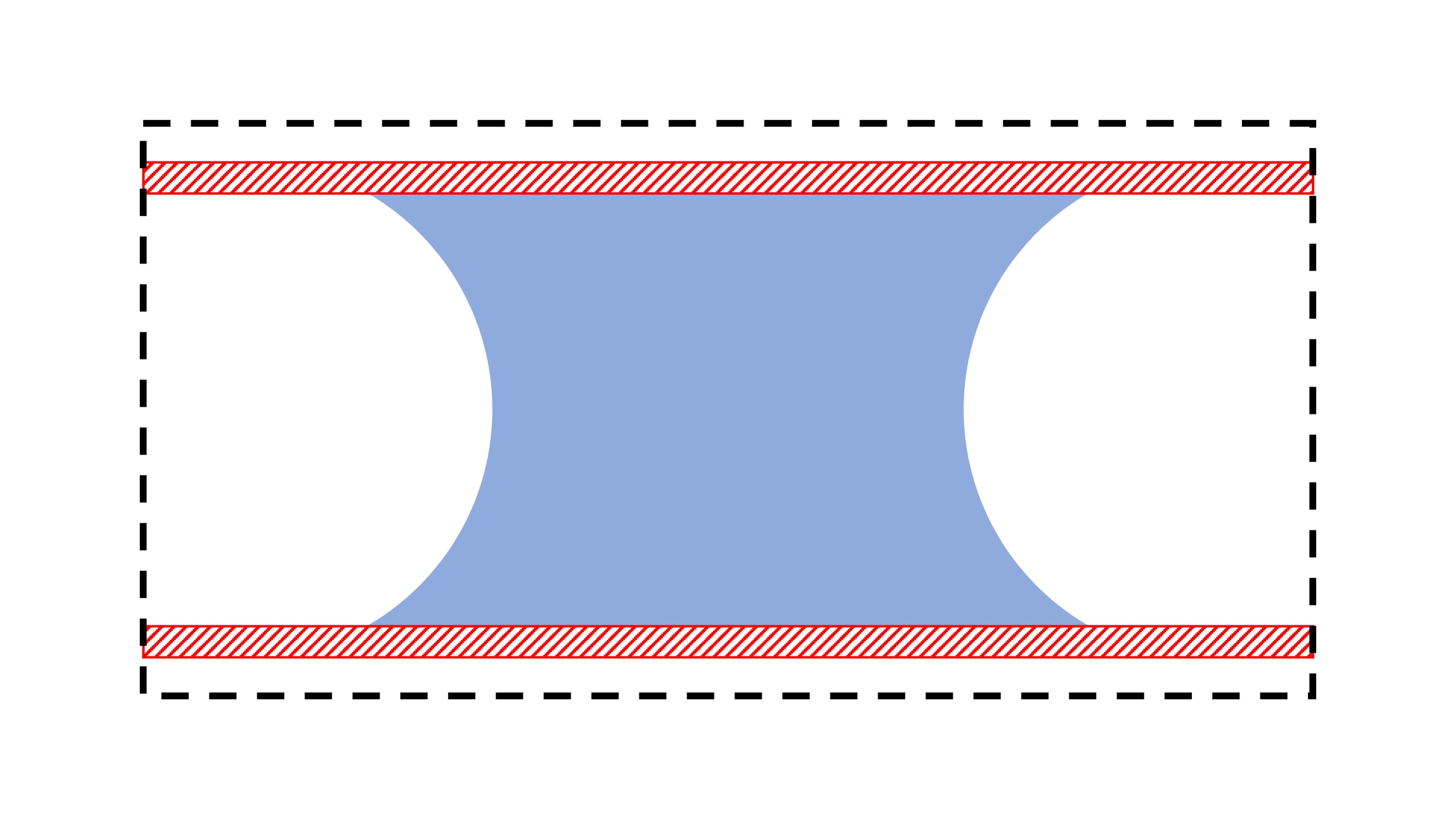}} & MENISCUS$\quad$ & Depletion layer & & $160\times4.7\times30.4$ \\
        \hline
    \end{tabularx}
    \caption{Type of systems simulated using Molecular Dynamics. The figures shown in the ``scheme" column outline the systems' geometry in the \textit{xz} plane. Dashed lines represent periodic boundary conditions, while solid lines represent purely repulsive walls preventing molecules from traveling across the boundaries perpendicular to the $z$ direction. The ``label" and ``goal" columns report the name with which the systems will be referred to in the article and the observables or transport coefficients that are to be obtained from simulations. The right-most column reports the reference sizes of simulation boxes, i.e. the sizes of initialized systems. $L^0_i$ may differ from the final edge length since equilibration at constant pressure necessarily changes the boxes' volume.}
    \label{tab:simulations}
\end{table}

In order to fully characterize the bulk, interfacial and wetting properties of the water-glycerol-silica combination, several different configurations of bulk liquid, liquid-vapor interfaces and liquid-vapor-solid contact lines need to be simulated. Table \ref{tab:simulations} offers a brief overview. Configurations of type SLAB are used to measure the liquid-vapor surface tension from the difference between lateral and normal components of pressure:
\begin{equation}
    \sigma = \frac{L_z}{2} \Big\{ P_{zz} - \frac{P_{xx}+P_{yy}}{2} \Big\} \; .
\end{equation}
Configurations of type DROP I and DROP II are used to obtain respectively the equilibrium and the dynamic contact angle. Droplets are prepared with a contact angle $\theta>\theta_0$ and simulated until fully spread. The initial contact angle is obtained by setting $q=-0.35$ $|e^-|$, which is then switched to $q=-0.72$ $|e^-|$ to simulate spontaneous wetting. Subsequently, the equilibrium contact angle is measured by fitting a circular cap to the liquid-vapor interface. Dynamic contact angle and contact line speed are measured by tracking the interface over time for 5 independent replicas of the same droplet spreading simulation with initial conditions sampled from a long equilibrium trajectory. Interface extraction is based on maps of liquid density binned on-the-fly while simulating and stored on structured rectangular grids; further details are presented in \ref{sec:interface}.

Configurations of type MENISCUS are used to extract information relative to near-wall liquid density layering. A meniscus configuration is more advantageous compared to a droplet configuration when extracting equilibrium interfacial information, as it allows simulating two independent liquid-solid interfaces simultaneously. 

\subsection{Shear viscosity}
\label{sec:visco-main}

The dispersion of values for the viscosity of SPC/E water reported in the literature is substantial, ranging from 0.64 cP \cite{hess2002viscosity} to 0.91 cP \cite{walther2013nanotube}. This hints towards an inherent difficulty in finding a robust procedure to calculate viscosity of liquids using Molecular Dynamics. Given the objective of our study, the measurement of shear viscosity deserves particular care. We utilize both equilibrium and non-equilibrium simulations and validate the former against the latter. Additional details on the techniques can be found in \ref{sec:viscosity}.

The equilibrium approach consists in simulating configurations of type BOX I at thermodynamic equilibrium for a sufficiently long time. The shear viscosity coefficient is then obtained by computing the integral of the off-diagonal components of the pressure tensor, that via the following Einstein relation:
\begin{equation}    \label{eq:linear-response}
    \eta = \lim_{t\rightarrow\infty}\frac{1}{6}\frac{V}{k_BT}\frac{d}{dt}\sum_{i\ne j} \expval{\Big(\int_0^tP_{ij}(\tau)d\tau\Big)^2} \; ,
\end{equation}
where angular brackets represent the ensemble average over several replicas. 

Conversely, the non-equilibrium approach consists in applying an external forcing to systems of type BOX II and measuring the response of the fluid. The forcing can have the form of a periodic acceleration field \cite{hess2002viscosity}:
\begin{equation}
    a_x(z) = \xi \cos(kz) \; , \quad k = \frac{2\pi}{L_z} \; .
\end{equation}
The fluid response is obtained directly by inspecting the flow profile in the direction parallel to the acceleration field, which can be shown to relax exponentially in time to:
\begin{equation}    \label{eq:non-equilibrium}
    u_x(z) = \frac{\xi\rho}{\eta k^2}\cos(kz) \; .
\end{equation}
In contrast to the equilibrium approach, the viscosity could in principle depend on the amplitude of the external acceleration. Therefore, a few values of $\xi$ need to be tried to assess the consistency of results.

\section{Simulation results}
\label{sec:analysis}

\subsection{Material parameters and transport coefficients}
\label{sec:analysis-coeff}

Table \ref{tab:coeff} reports the results obtained for the values of glycerol mass fraction under study. Liquid density is obtained from simulations of BOX I systems at constant temperature and hydrostatic pressure, that is after the box volume is allowed to relax to equilibrium. Density is found to increase with the mass fraction of glycerol, in quantitative agreement with literature (\cite{volk2018glydens}, see the supplementary information). 

Surface tension does not depend substantially on $\alpha_g$, implying a neutral effect of glycerol molecules on the liquid-vapor interface. A calculation of the accessible surface area of glycerol and water molecules at the liquid-vapor interface corroborates the absence of crowding or depletion of any of the two species. These results are partially is in contrast with experimental studies, where the surface tension of pure glycerol is reported to be about 10\% lower than the one of pure water \cite{takamura2012glycerol,yada2023prf}. We believe the reason for this discrepancy is the choice of water model, which underestimates the surface tension between pure water and vapour. We observe a slight improvement using TIP4P/2005, a more accurate and computationally expensive water model. In this case we observe a roughly 4\% drop between the surface tension of pure water and the one of pure glycerol. A sensitivity analysis on the estimate of contact line friction shows that using TIP4P/2005 instead of SPC/E would have a marginal effect on the scaling between contact line friction and viscosity and even the discrepancy with experimental results is not large enough to affect the conclusions of the study. The results of the sensitivity analysis are reported in the supplementary information.

For sufficiently low glycerol concentrations the equilibrium contact angle does not change significantly. Values for our guess of specific surface energy, i.e. partial charge on silica oxygens interacting with liquid molecules, produce contact angles that are in the range of the ones from the experimental work of Carlson et al. \cite{carlson2012contactline} and Yada et al. \cite{yada2023prf} for treated silica surfaces (e.g. silanized).

\begin{table}
    \centering
    \begin{tabular}{| l || l l l l |}
    \hline
    $\alpha_g$ & $\rho$ & $\sigma_{SPC/E}$ & $\sigma_{TIP4P/2005}$ & $\theta_0$ \\
      & [kg/m$^3$] & [$10^{-2}$Pa$\cdot$m] & [$10^{-2}$Pa$\cdot$m] & [$^\circ$] \\
    \hline
    \hline
    0.0 & 997.9$\pm$0.1 & 5.59$\pm$0.03 & 6.14$\pm$0.18 & 48.94$\pm$1.60 \\
    \hline
    0.2 & 1046$\pm$0.2 & 5.56$\pm$0.08 & 6.12$\pm$0.23 & 47.62$\pm$1.77 \\
    \hline
    0.4 & 1096$\pm$0.1 & 5.56$\pm$0.04 & 6.23$\pm$0.28 & 47.12$\pm$1.78 \\
    \hline
    0.6 & 1150$\pm$0.5 & 5.60$\pm$0.12 & 6.05$\pm$0.33 & 51.44$\pm$0.44 \\
    \hline
    0.8 & 1192$\pm$0.8 & 5.49$\pm$0.24 & 5.80$\pm$0.37 & 53.40$\pm$1.77 \\
    \hline
    1.0 & 1238$\pm$1.2 & 5.89$\pm$0.29$^{\dagger}$ & / & 76.06$\pm$0.47 \\
    \hline
    \end{tabular}
    \caption{Density, surface tension (for SPC/E and TIP4P/2005) and equilibrium contact angle over silica monolayers of liquid mixtures for the tested value of glycerol mass fraction. Density is simply computed from systems of type BOX I at equilibrium. The surface tension of pure glycerol ($\dagger$) is not reported for TIP4P/2005 since it does not depend on the water model.}
    \label{tab:coeff}
\end{table}

\begin{figure}[htbp]
    \centering
    \includegraphics[width=0.95\textwidth]{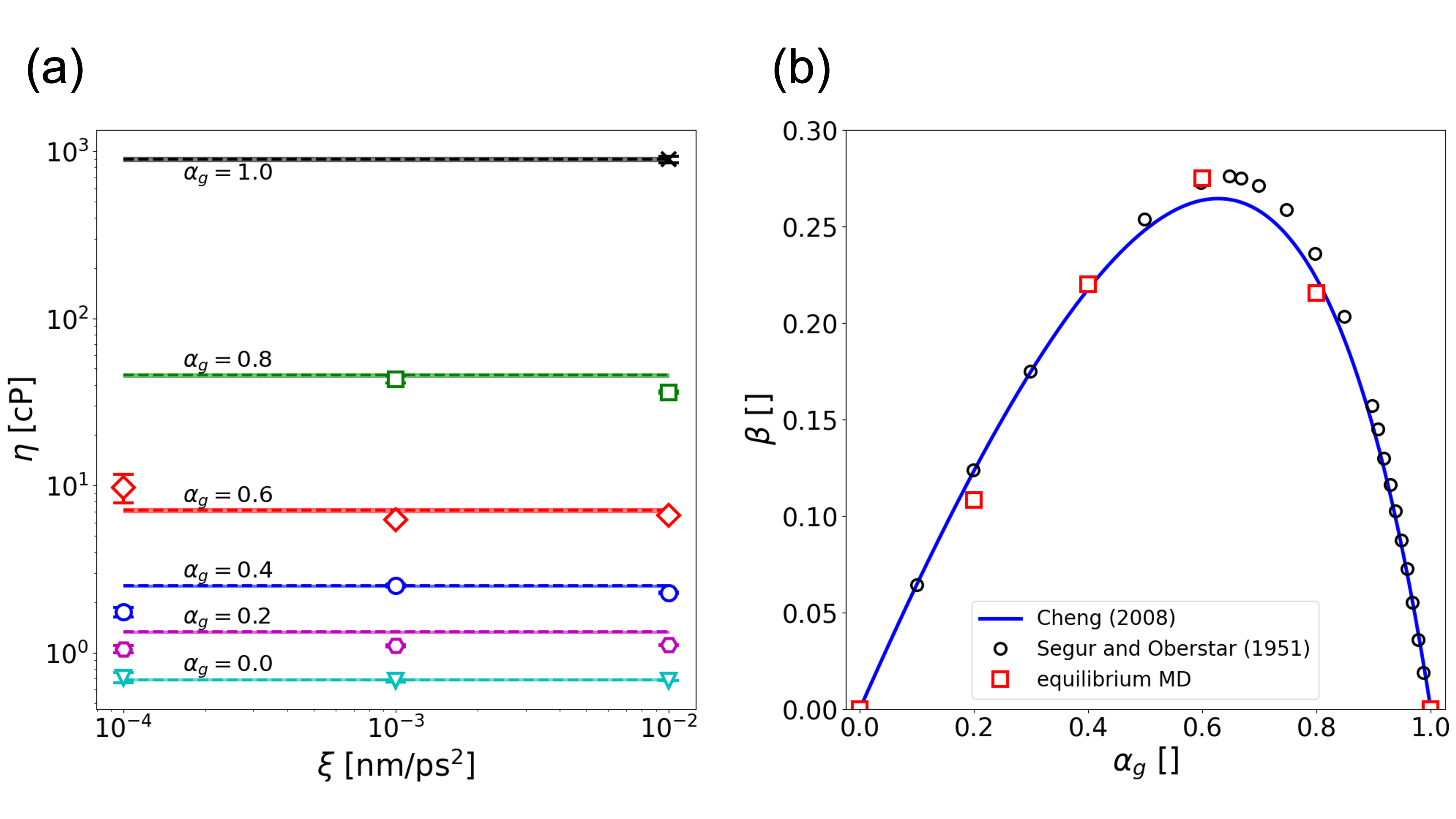}
    \caption{(a): Shear viscosity (in centiPoise) as a function of the non-equilibrium forcing amplitude $\xi$ and the mass fraction of glycerol $\alpha_g$. The markers with error bars indicate the results of non-equilibrium simulations, while dashed horizontal lines indicate the results of equilibrium simulations, i.e. Einstein relation; shaded areas represent the uncertainty of equilibrium results. (b): Comparison between Molecular Dynamics results, experimental results by Segur and Oberstar \cite{segur1951viscosity} and the best fit of the empirical model by Cheng \cite{cheng2008viscosity}; $\beta$ is defined in equation \ref{eq:cheng}.}
    \label{fig:viscosity-and-fit}
\end{figure}

The calculation of viscosity via equilibrium and non-equilibrium simulations is performed for each value of  $\alpha_g$. The results are presented in figure \ref{fig:viscosity-and-fit}a. Although it was not possible to obtain sufficiently converged results for large mass fractions of glycerol and small periodic perturbations, equilibrium and non-equilibrium approaches provide consistent results for all other combinations of $\alpha_g$ and $\xi$. In the following, the viscosity values obtained using Einstein's formula will be taken as reference, as they don't depend on any arbitrary forcing parameter. Furthermore, the simulation results are compared with the empirical model by Cheng \cite{cheng2008viscosity}:
\begin{equation}    \label{eq:cheng}
    \frac{\log(\eta/\eta_g)}{\log(\eta_w/\eta_g)} = (1-\alpha_g) + \beta(\alpha_g) \; , \quad \beta(\alpha_g) = \frac{ab\alpha_g(1-\alpha_g)}{a\alpha_g+b(1-\alpha_g)} \; ,
\end{equation}
$a$ and $b$ being fitting parameters, and with the experimental measurements by Segur and Oberstar \cite{segur1951viscosity}. Results are reported in figure \ref{fig:viscosity-and-fit}b and show quantitative agreement. This comparison constitutes evidence that the rheology of simulated aqueous glycerol solutions reflects physical reality.

\subsection{Contact line friction}
\label{sec:analysis-dynamic}

Contact line friction is estimated directly by fitting equation \ref{eq:linear-mkt} to the post-processed results of molecular simulations, with fitting parameters $\{\mu_f,\theta_0\}$ for each glycerol fraction. The equilibrium contact angle can be obtained from the intercept of the linear fit between $u_{cl}$ and $\cos\theta$, or can be imposed to the value obtained separately from equilibrium runs (table \ref{tab:coeff}). In order to incorporate both sources of information, a `lasso-like' regression is performed by imposing a weak constraint on the cosine of the contact angle, which corresponds to the minimization of the following functional:

\begin{equation}    \label{eq:lasso-fit}
    \mathcal{L}(\mu_f, \theta_0) = || \mu_f \vec{u}_{cl} - \sigma (\cos\theta_0-\cos\vec{\theta}) ||_2 + \zeta|\cos\theta_0-\cos\widehat{\theta}_0| \; ,
\end{equation}

being $\widehat{\theta}_0$ the value of the equilibrium contact angle reported in table \ref{tab:coeff}. The notation $\vec{\cdot}$ indicates the list of values obtained from molecular simulations. The contact line friction parameter resulting from the minimization of $\mathcal{L}$ is found to be insensitive to the penalty term; the penalty coefficient $\zeta$ is set to 0.1. Further information on uncertainty quantification can be found in the supplementary information.

\begin{figure}[htbp]
    \centering
    \includegraphics[width=0.95\textwidth]{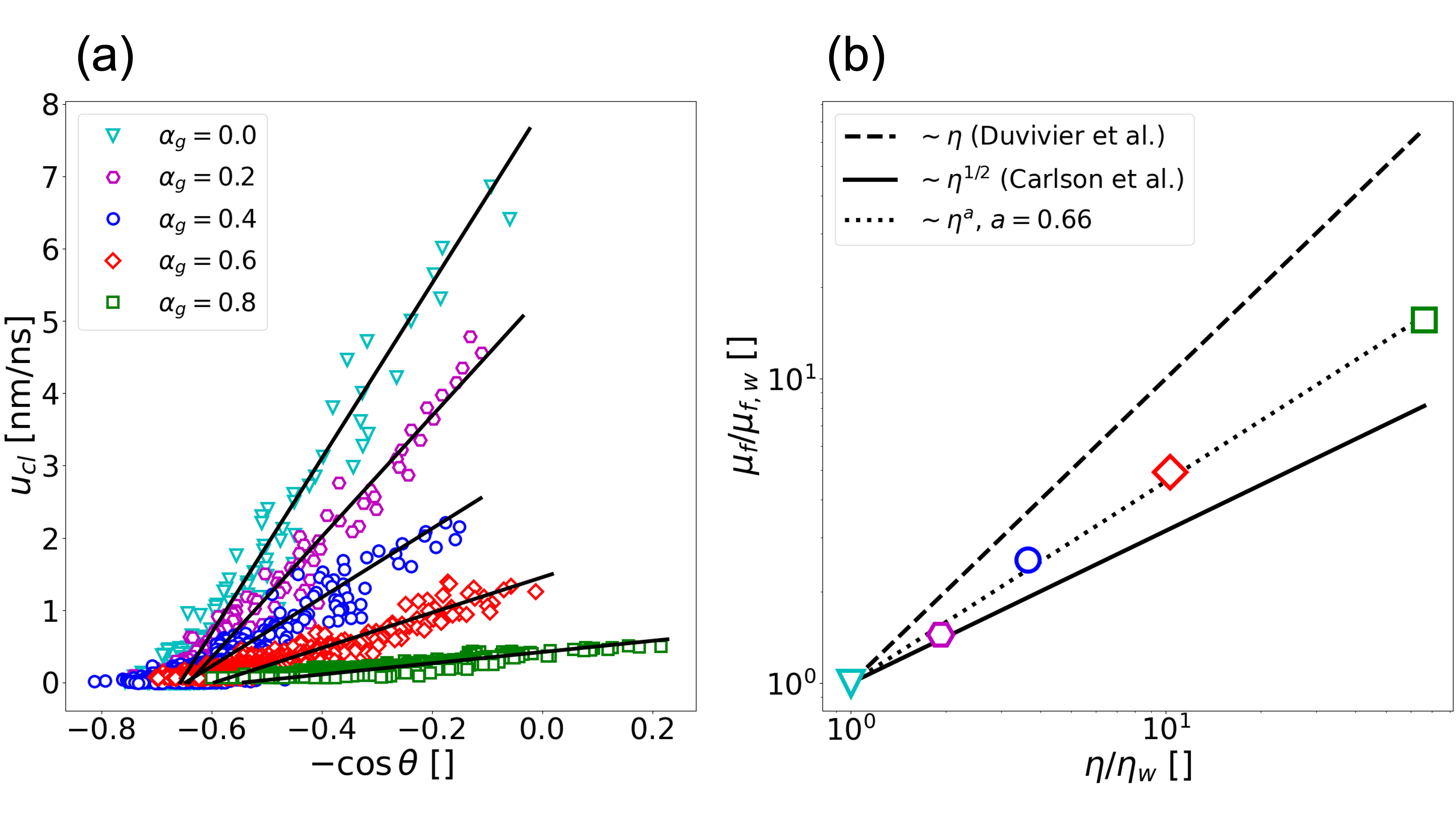}
    \caption{(a): Contact line speed against the opposite of the cosine of the dynamic contact angle. Markers represent measurements sampled from MD simulations results, whereas solid lines represent the best fit of equation \ref{eq:linear-mkt}, with cost function given by equation \ref{eq:lasso-fit}. (b): Log-log plot of the contact line friction coefficients against the shear viscosity coefficients, scaled on the reference value for water. The dotted line represents the best fit of MD results, while the solid and the dashed lines correspond respectively to the scaling laws found by Duvivier et al. \cite{duvivier2011experiment} and Carlson et al. \cite{carlson2012contactline}.}
    \label{fig:cl-friction-fit}
\end{figure}

\begin{table}
    \centering
    \begin{tabular}{ | l || l  l | }
        \hline
        $\alpha_g$ & $\eta$ & $\mu_f$ \\
        & [cP] & [cP] \\
        \hline
        \hline
        0.0 & 0.69$\pm$0.01 & 3.77$\pm$0.1 \\
        0.2 & 1.33$\pm$0.02 & 5.43$\pm$0.1 \\ 
        0.4 & 2.51$\pm$0.04 & 9.50$\pm$0.2 \\
        0.6 & 7.10$\pm$0.2 & 18.5$\pm$0.4 \\
        0.8 & 45.7$\pm$1.1 & 58.8$\pm$1.3 \\
        \hline
    \end{tabular}
    \caption{Bulk shear viscosity and contact line friction as a function of the mass fraction of glycerol. Note that contact line friction has the same physical units of viscosity.}
    \label{tab:viscosity_friction}
\end{table}

Figure \ref{fig:cl-friction-fit}a shows the measurements of the contact line speed against the dynamic contact angle, as well as the best fit of equation \ref{eq:lasso-fit}. Albeit much thermal noise remains even after averaging over a few replicas, a bijective relation between contact line friction and dynamic contact angle can clearly be observed for all values of $\alpha_g$: this is a signature of contact line friction. Furthermore, the linear MKT model appears suitable to capture the correlation between contact angle and contact line speed. Deviations from linear behavior due to advancing/receding asymmetry and high contact line speed have been previously reported \cite{toledano2020hidden,pellegrino2022asymmetry}; however, given that only advancing contact are simulated and that the spreading process is slow, especially for large glycerol concentrations, a linear model does suffice.

The values of the estimated contact line friction coefficients are reported in table \ref{tab:viscosity_friction}. Note that no result is reported for $\alpha_g=1.0$. On one hand, it was not possible to obtain a sufficiently wide range of contact line velocities from systems of type DROPLET II, given that the increase in viscosity caused the spreading rate to slow dramatically. On the other hand, interface tracking proved too noisy on smaller systems of type DROPLET I.

As shown in the log-log plot \ref{fig:cl-friction-fit}b, contact line friction does not scale linearly with viscosity: to the same relative increase in shear viscosity corresponds a \textit{smaller} increase in contact line friction. This can be also easily observed from table \ref{tab:viscosity_friction}. Figure  \ref{fig:cl-friction-fit}b reports the scaling laws found by Duvivier et al. and by Carlson et al. \cite{duvivier2011experiment,carlson2012contactline}; the scaling obtained from molecular simulations lies in between $\sim\eta$ and $\sim\eta^{0.5}$.

\subsection{Glycerol depletion of interfacial layers}
\label{sec:depletion}

\begin{figure}[htbp]
    \centering
    \includegraphics[width=0.95\textwidth,trim={0 0.5cm 0 1cm},clip]{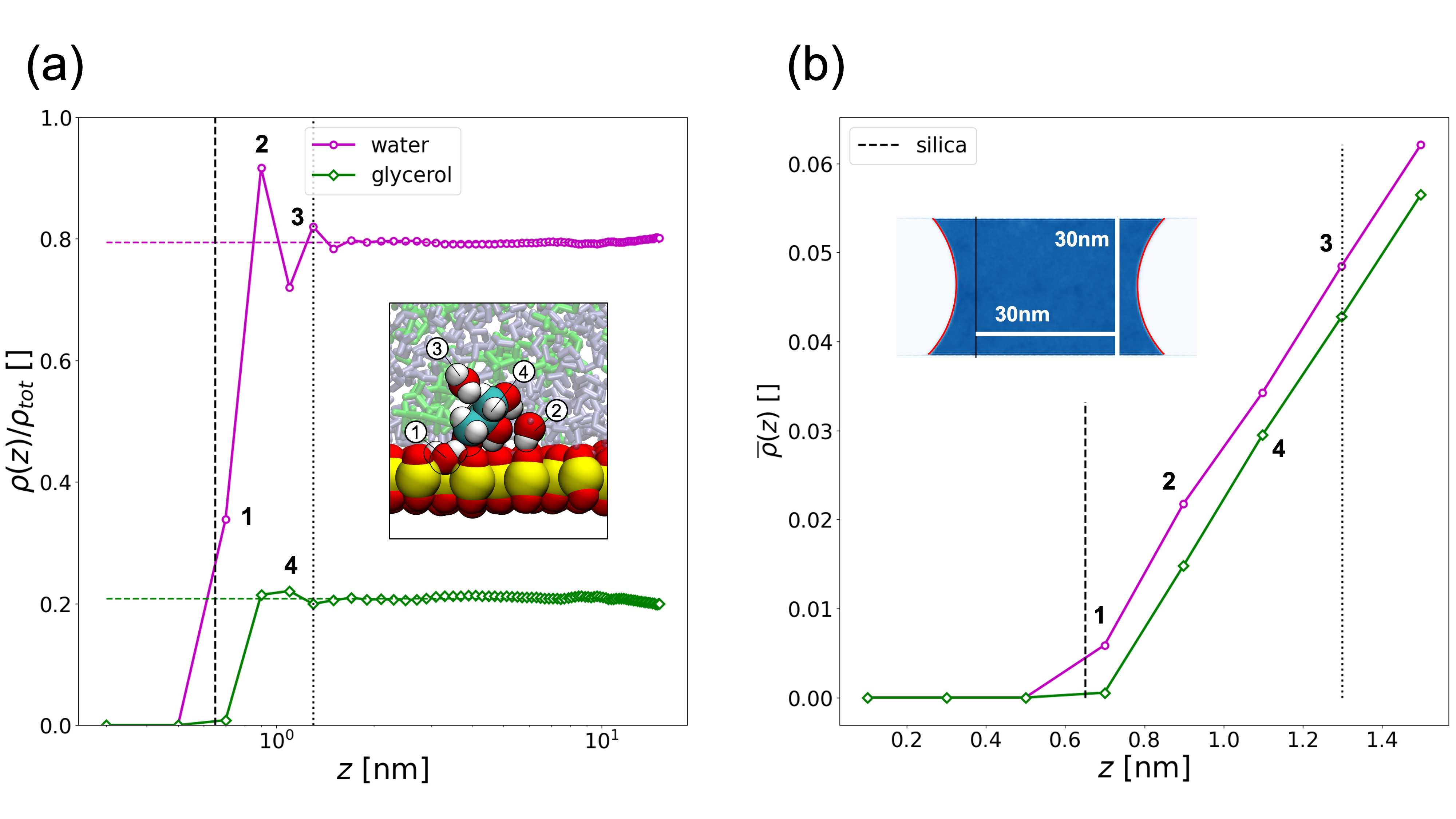}
    \caption{(a): Semi-logarithmic plot of the density profiles of water and glycerol as a function of the vertical coordinate, for the case $\alpha_g=0.2$. The density of the two components is re-scaled on the total average density in the bulk. The inset shows a snapshot of the near-wall position and orientation of water and glycerol molecules. (b) Cumulative density profile (equation \ref{eq:cumulative}) re-scaled on the total bulk density, showing depletion of glycerol molecules in the liquid-solid interfacial layer. The inset shows the section of the meniscus used to compute density profiles.}
    \label{fig:depletion}
\end{figure}

The last set of results we report in this section regards near-wall liquid density layering. For the sake of brevity, we will refer to the results of $\alpha_g=0.2$, but the same conclusions hold for all glycerol concentrations. Figure \ref{fig:depletion}a reports the normalized density profiles of water and glycerol along the vertical direction. Layering can be observed close to solid walls. Regarding water, 3 layers can be clearly spotted labeled 1, 2 and 3 in the figure: a layer of adsorbed molecules (1), a layer of molecules that are hydrogen-bonded with silica (2) and a third layer that bonds to molecules of layer 2 (3, indicated by a dotted line). Similar density oscillations have been observed experimentally for water wetting flat mica surfaces \cite{cheng2001mica}. As for glycerol, there appears to be no adsorbed layer; a layer with density slightly larger than the bulk can be observed at around 0.35nm from silicon atoms (4).

The spatial resolution of the density profile is limited by the cell size of the density binning grid, 0.2 nm. In order to detect whether the liquid-solid interface is crowded/depleted by a molecular species we compute the relative cumulative density along the vertical direction:
\begin{equation}    \label{eq:cumulative}
    \overline{\rho}(z) = \frac{\int_0^z\rho(\zeta)d\zeta}{\int_0^{L_z}\rho(\zeta)d\zeta} \; .
\end{equation}
The gap between cumulative profiles entails crowding of water in the liquid layers close to the interface, or conversely depletion of glycerol. In the next section we further discuss the implications of glycerol depletion and how it may be incorporated into MKT.

\section{Discussion}
\label{sec:discussion}
\subsection{Two-component contact line friction model}
\label{sec:multi-friction}
In real-world experiments, the viscosity of a wetting fluid can be tuned by changing its chemical composition, e.g. by the addition of a solute; it is reasonable to suppose that the effects of such a modification would not be limited to macroscopic rheology, but would also extend to microscopic kinetics. Furthermore, the molecular structure of a liquid in the proximity of interfaces may differ substantially compared to the one in the bulk: as shown in section \ref{sec:depletion}, the depletion of one species can be noticeable even in the case of a mixture of simple molecules wetting a flat surface. 

MKT has been originally formulated for the case of a single-component liquid: on the light of our results and of the considerations above it may be insufficient to explain wetting dynamics of solutions. An attempt to generalize MKT to multicomponent liquids has been proposed by Liang et al. \cite{liang2010multicomponent}. The model formulation stems from the decomposition of the total irreversible dissipation work at the contact line into the contribution from each molecular species. The contact line speed is therefore expressed as linear combination of the wetting speed of each component; in the limit of the molecular relaxation being much faster than the contact line speed, the linearized mobility relation reads:
\begin{equation}    \label{eq:multicomponent}
    u_{cl} = \sigma \sum_{i=1}^n \frac{\lambda^3_i\kappa^0_i}{\alpha_i^* k_BT} \big(\cos\theta_0-\cos\theta\big) \; ,
\end{equation}
being $n$ the number of molecular species. The superscript $\cdot^*$ denotes values that are localized at the liquid-wall interface, as opposed to the bulk. In the case of water-glycerol mixtures $n=2$. Instead of denoting glycerol and water with $i=\{1,2\}$ we will use the notation $\{g,w\}$; moreover $\alpha_g^*<\alpha_g$, owing to the results of section \ref{sec:depletion}. The dependence on viscosity is `hidden' in the equilibrium jump rate $\kappa_i^0\propto\kappa_{s,i}^0/(v_i\eta^*)$, where we highlight that the specific molecular volume will be species-dependent and that viscosity in the interfacial region may differ from the one in the bulk.

\subsection{Correction to linear scaling}
\label{sec:interf-visco}

\begin{table}
    \centering
    \begin{tabular}{ | l || l l | l l l l | }
        \hline
        $\alpha_g$ & $\expval{\tau^{hb}_w}$ & $\expval{\tau^{hb}_g}$ & $\lambda^3/v_w$ & $\lambda^3/v_g$ & $\alpha_g^*$ & $\eta^*$ \\
         & [ps] & [ps] & & & & [cP] \\
        \hline
        \hline
        0.2 & 2.38$\pm$0.40 & 2.03$\pm$0.37 & 5.23$\pm$0.03 & 1.07$\pm$0.02 & 0.186$\pm$0.022 & 1.147$\pm$0.080 \\  
        0.4 & 2.40$\pm$0.44 & 2.08$\pm$0.39 & 5.17$\pm$0.03 & 1.08$\pm$0.01 & 0.356$\pm$0.030 & 2.135$\pm$0.265 \\  
        0.6 & 2.50$\pm$0.40 & 2.17$\pm$0.35 & 5.07$\pm$0.05 & 1.08$\pm$0.01 & 0.502$\pm$0.032 & 4.334$\pm$0.754 \\  
        0.8 & 2.24$\pm$0.42 & 2.51$\pm$0.48 & 4.91$\pm$0.08 & 1.09$\pm$0.01 & 0.720$\pm$0.037 & 20.97$\pm$6.702 \\
        \hline
    \end{tabular}
    \caption{Average hydrogen bond lifetime of water and glycerol molecules with silica, alongside the parameters used to evaluate equation \ref{eq:corrected-friction}: the ratio between the cube of the lattice spacing and the per-molecule volume of water and glycerol, the local mass fraction of glycerol and the estimated local viscosity.}
    \label{tab:interfacial_friction}
\end{table}

\begin{figure}[htbp]
    \centering
    \includegraphics[width=0.95\textwidth,trim={0 0 0 0},clip]{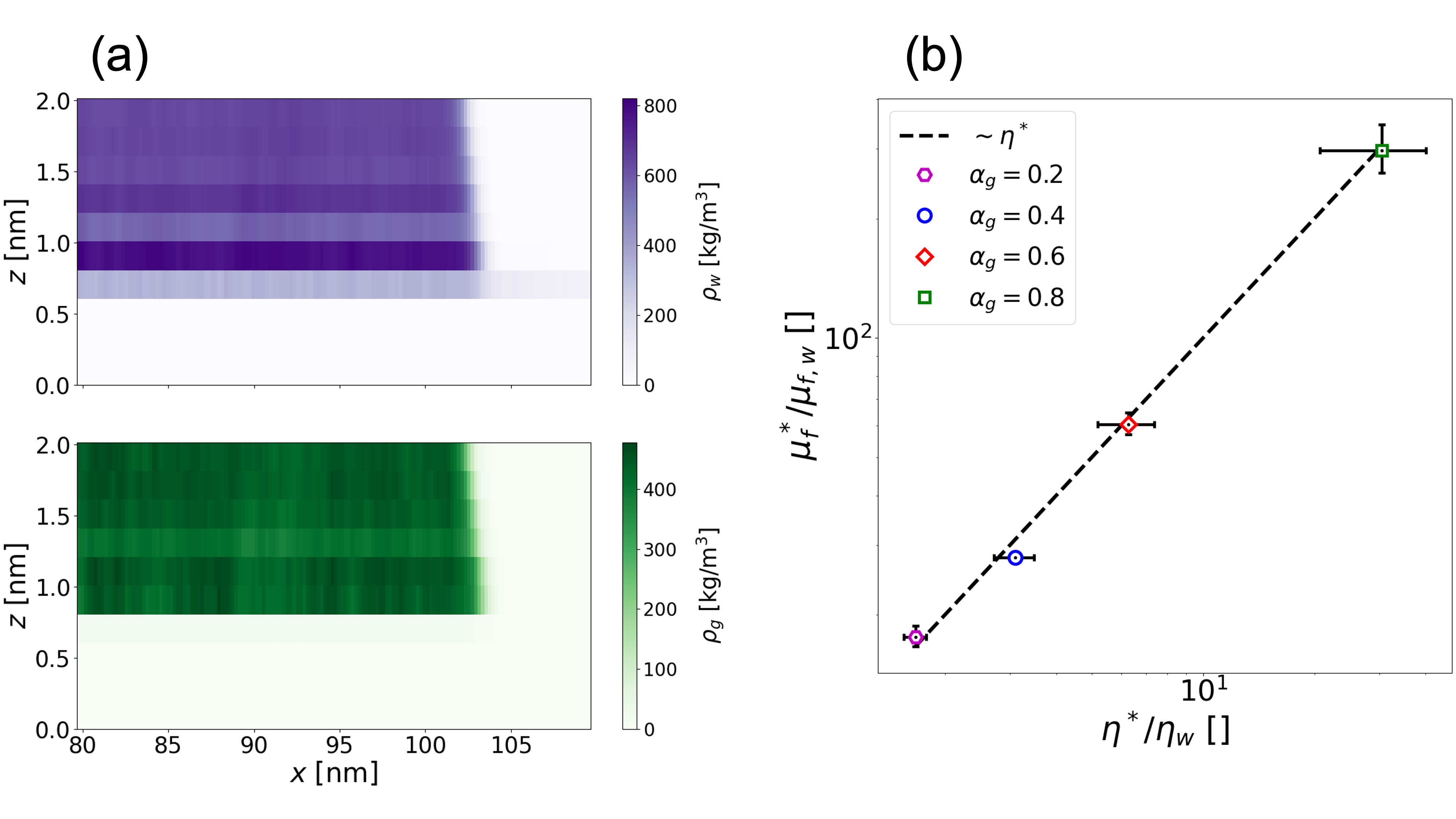}
    \caption{(a) Zoom on the lower-right contact line of a MENISCUS system showing the density map of water (purple) and glycerol (green). (b) Log-log plot of the corrected contact line friction (scaled on the friction of water) against the corrected interfacial viscosity (scaled on the viscosity of water), showing linear scaling between the two coefficients. Error bars correspond to the uncertainty (one standard deviation) in $\alpha_g^*$, see supplementary information.}
    \label{fig:adjusted-scaling}
\end{figure}

We now discuss how to simplify equation \ref{eq:multicomponent} and obtain a correction to the relation between contact line friction and viscosity. We consider the depletion of glycerol in the interfacial region and the difference in excluded volume between the two molecular species to be the primary factors to correct for. First, we assume that the layering and depletion effects discussed in section \ref{sec:depletion} propagate up to the contact line. This assumption is substantiated by visual inspection of density maps (figure \ref{fig:adjusted-scaling}a). There are strong indications that the main contribution to contact line displacement is provided by the jumping motion of molecules that are temporarily hydrogen bonded to the substrate \cite{johansson2019friction,pellegrino2022asymmetry}. Therefore, we assume that $\lambda_w\simeq\lambda_g\simeq d_{hex}=0.45$ nm, since the spacing of the hexagonal silica quadrupole lattice determines where the hydrogen bonding sites are located. We also assume that the equilibrium jump rate on the substrate does not change substantially with the addition of glycerol. This assumption is substantiated by the following observations: on the one hand, molecules interacting with the surface spend most their time bonded with surface molecules rather than in the transition state between jumps; on the other hand, the typical lifetime of hydrogen bonds between water and silica and between glycerol and silica does not depend substantially on glycerol concentration.

Owing to the assumptions discussed above, we can simplify equation \ref{eq:multicomponent} and obtain a new relation between a re-scaled friction coefficient and viscosity:
\begin{equation}    \label{eq:corrected-friction}
    \mu_f^* = \Big[\frac{\lambda^3}{\alpha_g^*v_g}+\frac{\lambda^3}{(1-\alpha_g^*)v_w}\Big]\mu_f \propto \frac{k_BT}{\kappa^0_s}\eta^*
\end{equation}
We now introduce a further assumption: $\eta^*=\eta(\alpha_g^*)$, i.e. the change of interfacial viscosity is only due to the depletion of one molecular species. This assumption is consistent with the \textit{local average density model} (LADM), whereby the local shear viscosity is identified with the viscosity of the homogeneous fluid at the local average density, or in this case at the average relative concentration \cite{bitsanis1987ladm}. We remark that this may be a very blunt simplification. It is known that the addition of glycerol disrupt the local structure of the hydrogen bonds network of bulk-like water \cite{dashnau2006glycerol}, causing viscosity to deviate from ideal solution theory in its dependence on glycerol concentration. However, it is not clear how this effect would influence viscosity in nanoconfinement, i.e. in the occurrence of depletion and adsorption. Layering in nanoconfinement is known to greatly affect the local configurational shear viscosity close to solid walls in single-component liquids, requiring a more refined model then LADM \cite{hoang2012confined}. Hence, assuming that the depletion, i.e. local change of mass fraction, is the only factor affecting local viscosity should be considered a `zero-order' approximation.

We can utilize the fit of equation \ref{eq:cheng} to re-compute viscosity with mass fraction values at the liquid-solid interface. The molecular volumes $v_w$ and $v_g$ are measured indirectly from the free volume and the volume occupied by each species in systems of type BOX I, while $\alpha_g^*$ is simply computed by counting molecules within 0.8 nm from the solid wall in systems of type MENISCUS (dotted line of figure \ref{fig:depletion}b). Provided the interfacial viscosity, the local mass fraction and the molecular volumes, expression \ref{eq:corrected-friction} can be evaluated. The values of the coefficients used to evaluate equation \ref{eq:corrected-friction} are reported in table \ref{tab:interfacial_friction}. Appendix \ref{sec:near-wall-friction} briefly explains how the coefficients are obtained from molecular simulations.

Values for the depletion-corrected friction coefficient and the interfacial viscosity are presented in figure \ref{fig:adjusted-scaling}b. Linear scaling between the two coefficients is re-obtained. A leading-order correction for depletion and excluded volume effects is therefore enough to reconcile the results of Molecular Dynamics simulation with Molecular Kinetic Theory.

\subsection{Comparison with experimental studies}

As mentioned in the introduction, some dynamic contact angle experiments have found a linear correlation between contact line friction and viscosity. In this section we address the source of the mismatch between experiments and molecular simulations. It is very possible that in some cases contact line friction would be mainly determined by liquid-liquid interactions rather than liquid-surface indications. These are cases of \textit{viscous coalescence analogy} \cite{doquang2015wetting}, occurring for example when a particularly viscous and homogeneous liquid droplet spreads on a hydrophobic or slippery surface. In such cases it is not unreasonable to expect that experiments would find a 1:1 correlation between viscosity and friction. Regardless, the method to estimate the contact line friction coefficient can by itself `screen' molecular-scale effects with viscous interface bending. If the coefficient is estimated by applying equations \ref{eq:linear-mkt} to a optically measured contact angle, then what the experiment is effectively measuring is the proportionality between spreading speed and the difference between the equilibrium and an apparent contact angle:
\begin{equation}    \label{eq:apparent-mkt}
    \mu_f^{app}u_{cl} = \gamma\big(\cos\theta_0-\cos\theta_{app}\big) \; .
\end{equation}
By using Petrov-Petrov combined molecular-hydrodynamic expression for the apparent contact angle of an advancing contact line, it is possible to roughly estimate the relation between $\mu_f^{app}$ and $\mu_f$ \cite{petrov1992combined}:
\begin{equation}    \label{eq:petrovx2}
    \theta_{app}^3 \simeq \theta^3(u_{cl};\mu_f) + \frac{\eta u_{cl}}{\gamma}\log\big(z_{obs}/l^*\big) \; ,
\end{equation}
where $z_{obs}$ is the distance from the contact line where the contact angle is measured and $l^*$ is a microscopic length scale, which can be interpreted as the lower-scale cutoff for the interface curvature. We can substitute equation \ref{eq:linear-mkt} into \ref{eq:petrovx2} for the relation $\theta(u_{cl})$ and \ref{eq:petrovx2} into \ref{eq:apparent-mkt} for the apparent contact angle. After linearizing for small wetting speeds we are left with the following expression:
\begin{equation}    \label{eq:apparent-friction}
    \mu_f^{app} \simeq \mu_f + \frac{3\log\big(z_{obs}/l^*\big)}{\theta_0}\eta \; .
\end{equation}
Two considerations can be drawn from equation \ref{eq:apparent-friction}: (i) If the contact angle is measured sufficiently far from the contact line, then it is only reasonable to observe a linear scaling between apparent contact line friction and viscosity. (ii) The apparent contact line friction always over-estimates molecular friction; the effect is exacerbated the further the measurement of the contact angle is from the contact line,  the more viscous the spreading liquid is and the more hydrophilic the surface is. 

This last point helps explaining why contact line friction coefficients measured in molecular simulations are usually an order of magnitude smaller than the ones reported in experimental studies using optical microscopy to measure the dynamic contact angle. However, even when the dynamic contact angle is treated as a hidden parameter and contact line friction is inferred from spreading rates, the reported friction coefficient might still be substantially larger the typical values obtained from simulations \cite{carlson2012universal,carlson2012contactline,li2023friction,yada2023prf}. We believe this is due to the unavoidable presence of surface defects such as topographic roughness and chemical heterogeneity, which have a great impact on contact line mobility \cite{joanny1984hysteresis,perrin2016defects}. Surface defects are rarely studied with molecular simulations of wetting dynamics since the scale of defects is sometimes much larger than what can be simulated and because reproducing realistic nanoscale defects often requires more advanced molecular modeling.

\subsection{Implications for the wetting properties of mixtures and solutions}

Heterogeneous complex liquids are known to exhibit an interesting rheology. For example, fluids containing colloidal particles or polymers (\textit{nanofluids}) have been studied because of their peculiar wetting properties. Self-assembly close to solid surfaces and migration away/towards high-shear regions can either reduce or increase spreading friction \cite{lu2016review}. In other cases, the addition of polymers can affect bulk rheology (e.g. by inducing shear thinning and viscoelastic behaviors) without affecting contact line friction \cite{yada2023prf}. Our results indicate that non-trivial correlations between bulk flow and wetting can also occur in the case of simple liquids when they are formed by multiple molecular species.

On high-friction surfaces (i.e. in cases where $\mu_f\gg\eta$, where wetting is thus \textit{substrate-dominated} \cite{doquang2015wetting}), the preferential affinity of substrate molecules for one of the liquid species can lead to a weak modulation of contact line friction upon changes in concentrations, whereas viscosity can be affected much more substantially. It is not unreasonable to also envision the opposite scenario, whereby the addition of molecular species with high affinity with the substrate would increase contact line friction without producing any dramatic change of bulk rheology. Such behaviors could be desirable in applications where the time scale of wetting is shorter than the possibly long relaxation times of polymers or diffusion times of colloidal particles, such as droplet impact or rapid capillary rise.

\section{Conclusions}
\label{sec:conclusions}

We have presented the results of Molecular Dynamics simulations aimed to investigate the correlation between contact line friction and viscosity of water-glycerol droplets spreading over flat silica-like surfaces. The shear viscosity coefficient has been computed from equilibrium and non-equilibrium simulations and has been found to reflect the rheology of real water-glycerol solutions. The contact line friction coefficient has been obtained directly by fitting linearized Molecular Kinetic Theory to the contact angle and the contact line speed measured from simulations. The scaling between contact line friction and viscosity has been determined to be sub-linear, contrarily to the predictions of single-specie MKT. A correction inspired by multicomponent MKT has been proposed and linear scaling between interfacial friction and viscosity has been recovered by accounting for the local change in mass fraction and the difference of excluded volume between water and glycerol.

Even if the effect of molecular depletion can be noticeable on flat substrates, simulating wetting dynamics on idealized surfaces constitutes a significant simplification. In the case of aqueous glycerol droplets, topographical roughness is known to affect the equilibrium contact angle on silica surfaces \cite{hitchcock1981experiment,moldovan2014experiment}. Therefore, simulating droplet spreading on more realistic surfaces is a natural continuation of this work. Furthermore, the assumptions introduced in section \ref{sec:interf-visco} to correct the scaling between shear and friction coefficients may be relaxed to address cases where to a change in viscosity corresponds a substantial change in molecular kinetic rates, such as when the wetting fluid or the surface are cooled down or heated up.

The most significant conclusion of this work is that molecular-scale effects, which cannot be easily captured in a continuous description of liquid-solid interfaces, contribute substantially to contact line friction. Even the rather simple case of an aqueous solution wetting a flat surface shows deviation from the postulated linear scaling. In essence, molecular heterogeneity emerges at the interface scale and hence needs to be accounted for in order to correctly characterize wetting phenomena.

\section*{Supplementary information}
See supplementary information at [URL will be inserted by publisher] for additional details on uncertainty quantification, the measurement of molecular interfacial observables and the calculation of shear viscosity.

\section*{Data availability statement}
All input of molecular simulations are available on Zenodo \cite{zenodo_wg_surftens,zenodo_wg_viscoeq,zenodo_wg_viscone,zenodo_wg_caeq,zenodo_wg_cane,zenodo_wg_mesc}. Part of the output of is also available; very heavy files such as fully-detailed trajectory have been excluded. Zenodo datasets also contain simple Python scripts to open and visualize the output of molecular simulations. More advanced analysis scripts can be provided under reasonable request. The Gromacs fork for the density binning code is available on the Github repository curated by Dr. Petter Johansson \cite{binningcode}.

\section*{Acknowledgements}
We wish to thank Prof. Shervin Bagheri at KTH, Prof. Gustav Amberg at Södertörns högskola and Dr. Petter Johansson at Université de Pau for the useful feedback. Numerical simulations were performed on resources provided by the Swedish National Infrastructure for Computing (NAISS snic2022-1-15) at PDC, Stockholm. We acknowledge funding from Swedish Research Council (INTERFACE Centre grant No. VR-2016–06119 and grant No. VR-2014–5680).

\section*{Author contributions}
\textbf{Michele Pellegrino}: conceptualization (equal), project administration (equal), data curation, formal analysis, investigation, methodology, visualization, writing - original draft, writing - review \& editing (equal). \textbf{Berk Hess}: conceptualization (equal), project administration (equal), supervision, writing - review \& editing (equal).

\appendix

\section{Molecular Dynamics simulation}
\label{sec:md}

The parametrization of molecular simulations is analogous to the one of \cite{lacis2021pellegrino} and \cite{pellegrino2022asymmetry}. We report the information needed to reproduce the simulations. Configuration files can be found on Zenodo \cite{zenodo_wg_surftens,zenodo_wg_viscoeq,zenodo_wg_viscone,zenodo_wg_caeq,zenodo_wg_cane,zenodo_wg_mesc}.

\begin{table}
\begin{center}
\begin{tabular}{llll}
    Oxygen mass & \quad$m_O$ & \quad$9.95140$ & u \\
    L-J well depth (silicon) & \quad$\varepsilon_{Si}$ & \quad$0.2$ & kJ mol$^{-1}$ \\
    L-J char. distance (silicon)  & \quad$\sigma_{Si}$ & \quad$0.45$ & nm   \\
    L-J well depth (oxygen) & \quad$\varepsilon_{O}$ & \quad$0.65019$ & kJ mol$^{-1}$ \\
    L-J char. distance (oxygen) & \quad$\sigma_{O}$ & \quad$0.316557$ & nm \\
    Si-O bond distance & \quad$d_{so}$ & \quad$0.151$ & nm \\
    Hexagonal lattice spacing & \quad$d_{hex}$ & \quad$0.45$ & nm \\
    Restraint force constant & \quad$\kappa_O$ & \quad$10^5$ & kJ mol$^{-1}$ nm$^{-2}$ \\
    Partial charge (oxygen) & \quad$q$ & \quad$-0.72$ & $|\mbox{e}^-|$
\end{tabular}
\end{center}
\caption{Parametrization of the force field of silica quadrupoles from \cite{lacis2021pellegrino}.}
\label{tab:MD-silica-char}
\end{table}

The force field of silica quadrupoles can be decomposed into a short-ranged non-bonded van der Waals term, a long-range attractive electrostatic term and a restrain term. Short-range non-bonded interactions are computed via the Lennard-Jones potential $U_{LJ}(r)=4\varepsilon[(\sigma/r)^{12}-(\sigma/r)^6]$ and long range electrostatic via the Coulomb potential $U_C(r)=(q_iq_j)/(4\pi\epsilon_0 r)$. Wall atoms are restrained to fixed positions in space with a spring potential of the kind: $U_R(r)=k(r-r_{ref})^2/2$. Restraining surface atoms instead of freezing them in place allows momentum exchange between liquid and solid molecules, which is necessary in order to avoid very large and unphysical slip lengths \cite{sega2013slipsubstrate}. Covalent bonds between oxygen and silicon atoms are treated as rigid constraints. Silicon atoms are treated as virtual sites without mass. It has been shown that the wetting behaviour on this type of surface is qualitatively different from the one of simple Lennard-Jones crystals and it is better suited to explore wetting dynamics on hydrophilic surfaces \cite{johansson2015physicochemistry}. Table \ref{tab:MD-silica-char} reports the force field parameters for silica.

Lennard-Jones parameters for the interactions between silica and water or glycerol are generated via the geometric combination rule. Short-range non-bonded interactions are computed using a Verlet list and 1 nm cutoff distance in real space. Electrostatic interactions are computed using the Particle Mesh Ewald method (PME), with 1 nm cutoff distance for the calculation in real space and 0.15 nm grid spacing for the calculation in the reciprocal space.

Integration over time is performed using the leapfrog algorithm with a time step of 2 fs. Systems are thermalized using the Bussi-Donadio-Parrinello thermostat (\texttt{v-rescale} in Gromacs), with a 1 ps coupling time. Ideal purely-repulsive Lennard-Jones walls are placed beyond the silica surfaces at the location of periodic boundary conditions along the vertical direction ensuring no interaction between periodic images along $z$.

\section{Interface extraction}
\label{sec:interface}

The measurement of the contact line speed and contact angle requires samples of the position of the liquid-vapor interface over time. Interface extraction for configurations of type DROP I and DROP II follows the procedure already presented in \cite{lacis2021pellegrino} and \cite{pellegrino2022asymmetry}. 

Two-dimensional ($xz$) density maps for the liquid-vapor systems are computed by binning the atomic positions on-the-fly, that is concurrently with the running simulation, on a structured grid with spacing $\Delta x=\Delta z\simeq0.2$ nm. The position of the interface along the vertical direction is in turn obtained by linearly interpolating the center of the two cells where the density is closer to half the bulk liquid density.

The contact angle is obtained in two different ways depending on the configuration and the goal. When running configurations of type DROP I the contact angle is identified via the tangent to the best least-squares circle fit of the interface shape obtained after a sufficiently long time. On the other hand, when simulating configurations of type DROP II we are interested in the non-equilibrium transient; in this case the contact angle is obtained by linearly interpolating the first 3 interface points, starting from 0.4 nm above the reference position of silicon atoms.

\begin{figure}[htbp]
    \centering
    \includegraphics[width=0.95\textwidth,trim={2.25cm 0.75cm 1.75cm 1.25cm},clip]{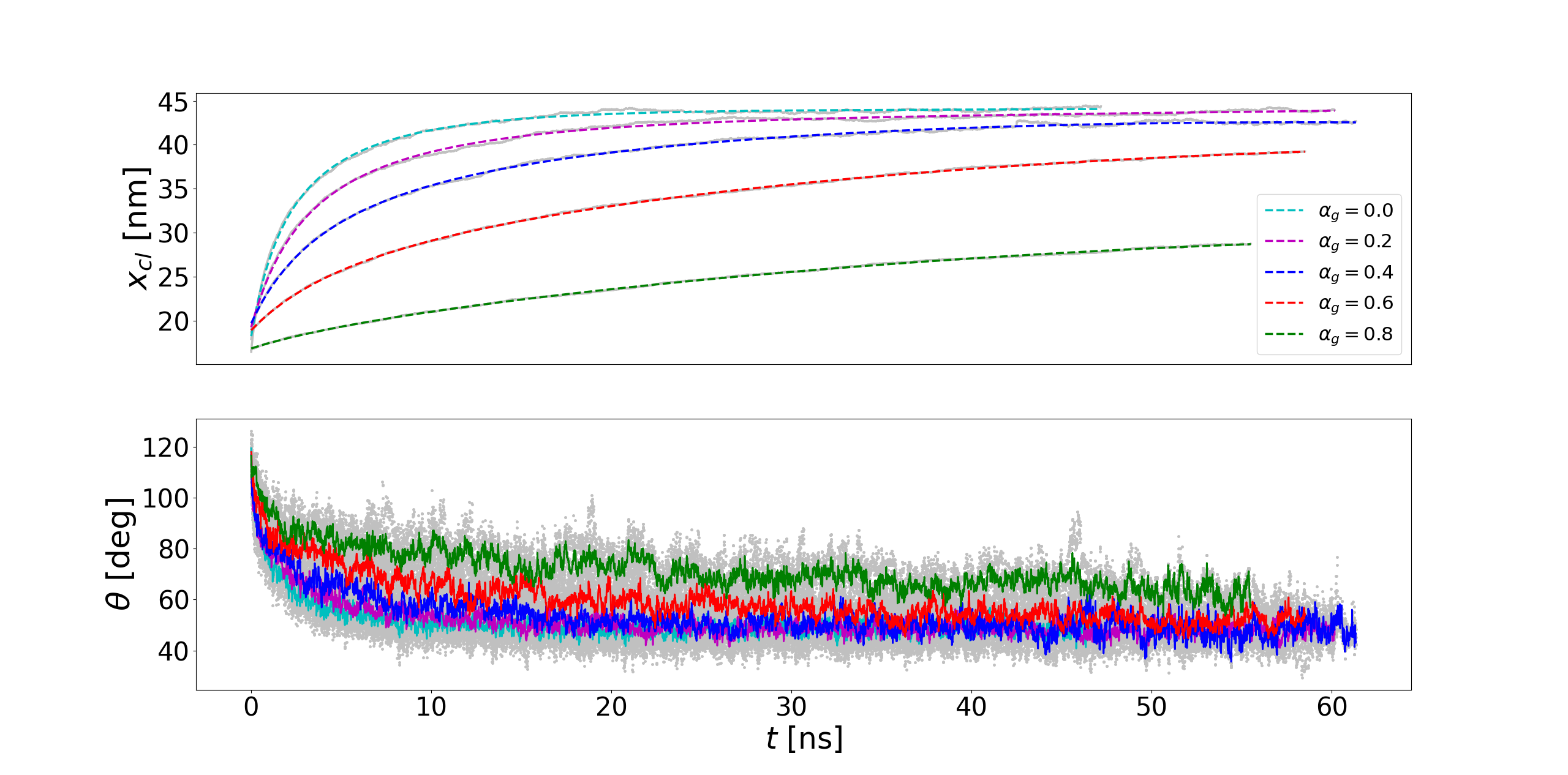}
    \caption{Relaxation of the contact line position and the dynamic contact angle over time. Gray lines and dots represent the results extracted directly from molecular simulations without post-processing (fitting for the contact line position and replica averaging for the contact angle).}
    \label{fig:relaxation-postproc}
\end{figure}

The speed of the contact line is obtained from the displacement of the contact line from its position at the beginning of the simulation. Due to thermal fluctuations, the position of the contact line is not differentiable over time; hence it is first fitted with a rational polynomial:

\begin{equation}
    x_{cl}(t) = \frac{a_3 t^3 + a_2 t^2 + a_1 t + a_0}{a_{-1} t + 1} \; ,
\end{equation}

and then differentiated analytically \cite{toledano2020hidden}. Constraints on the coefficients $a_j$ are imposed to ensure that the velocity is always positive and that the acceleration is always negative, as expected in an overdamped spreading regime with no inertial oscillations. On the other hand, no additional filtering is performed on the signal of the contact angle over time except replica averaging. Figure \ref{fig:relaxation-postproc} shows the relaxation curves of the contact line position and the dynamic contact angle over time for the configurations of type DROP II.

\section{Measurement of shear viscosity}
\label{sec:viscosity}

In this appendix we briefly describe how viscosity is computed from molecular simulations. A thorough explanation of equilibrium and non-equilibrium methods for the calculation of transport coefficients is outside the scope of this article. Nonetheless, this appendix, the additional details reported in the supplementary information and the data in \cite{zenodo_wg_surftens,zenodo_wg_viscoeq,zenodo_wg_viscone,zenodo_wg_caeq,zenodo_wg_cane,zenodo_wg_mesc} ensure the reproducibility of our calculations.

We first illustrate how viscosity is obtained using the equilibrium approach. Systems of type BOX I are first equilibrated at constant temperature (300 K) and pressure (1 bar) for 4 ns, after which long simulations ($>$100 ns) are performed at constant temperature and volume. The ensemble average is performed over 60 replicas, being either parallel runs with different initial conditions sampled from the equilibration run or different slices of the same molecular dynamics trajectory. The viscosity coefficient is obtained by linear regression of the ensemble average against time. Given the diffusive nature of the observable, the regression is weighted on the estimated standard error at a given time \cite{zhang2015viscosity}. The regression is performed within a time window determined on one hand by convergence to linear behavior (visually inspected) and on the other by a 10\% threshold on the standard error. 

The viscosity coefficient obtained using the non-equilibrium approach is computed by fitting the velocity field binned from molecular trajectory to expression \ref{eq:non-equilibrium}. As briefly explained in section \ref{sec:visco-main}, a drawback of this approach is that a few values of the intensity of the external field $\xi$ need to be screened when comparing the resulting viscosity with the one obtained with the equilibrium approach. In principle, a very small value of $\xi$ may be desirable in order to remain consistent with the `linear response' assumption. However, the signal-to-noise ratio on the flow field is better the larger the intensity of the perturbation. We considered 3 logarithmically spaced values for $\xi=\{10^{-2}, 10^{-3}, 10^{-4}\}$ nm/ps$^2$. Configurations of type BOX II equilibrated for 4 ns at constant temperature and pressure. The non-equilibrium simulations are then performed at constant volume and last at least 4 ns or until less than 10\% relative error (computed via bootstrapping) is achieved on the estimated viscosity.

\section{Interfacial friction}
\label{sec:near-wall-friction}

In this last appendix we report additional information on the coefficients reported in table \ref{tab:interfacial_friction}, used to evaluate equation \ref{eq:corrected-friction}. The lifetime of hydrogen bonds between liquid and solid molecules is computed using the Gromacs command \texttt{gmx hbond -life} on 3 ns long trajectories sampled every 1 ps. The per-molecule volume of water and glycerol can be computed indirectly by running \texttt{gmx freevolume} and supplying to the \texttt{-select} flag the group name of water or glycerol molecules. The per-molecule volume is then computed knowing the total number of molecules of one species and the difference in excluded volume between the other species and the total system. The local mass fraction of glycerol is computed in the layering region of MENISCUS systems within 0.8 nm from the silica wall and sufficiently far away from the liquid-vapor interface (as shown in figure \ref{fig:depletion}) using on-the-fly binned density maps. The effective local viscosity is computed by using equation \ref{eq:cheng} with $\alpha^*_g$ instead of $\alpha_g$.

\bibliography{apssamp}

\end{document}



\title{{\normalsize\normalfont Near-wall depletion and layering affect contact line friction of multicomponent liquids}\\Supplementary Information}

\author{M.~Pellegrino}
\author{B.~Hess}
\affiliation{Swedish e-Science Research Centre, Science for Life Laboratory,
Department of Applied Physics KTH, 100 44 Stockholm, Sweden}

\maketitle

\section{Material parameters}  \label{app:material-parameters}

In this section we report the results of the calculation of some material and kinetic parameters which substantiate the finding of the main article. We also report the Gromacs commands used to compute said parameters.

The density profile of aqueous glycerol solutions along one direction can be computed from configurations of type BOX (table 1 in the main article) by simply running \texttt{gmx density} and supplying the molecular trajectory (\texttt{.trr} file, \texttt{-f} flag) and the binary topology file (\texttt{.tpr} file, \texttt{-s} flag). The average density is then obtained simply by running \texttt{gmx analyze -av} and supplying the \texttt{density.xvg} file to the \texttt{-f} flag.

\begin{figure}[htbp]
    \centering
    \includegraphics[width=1\textwidth,trim={0 0 0 0}, clip]{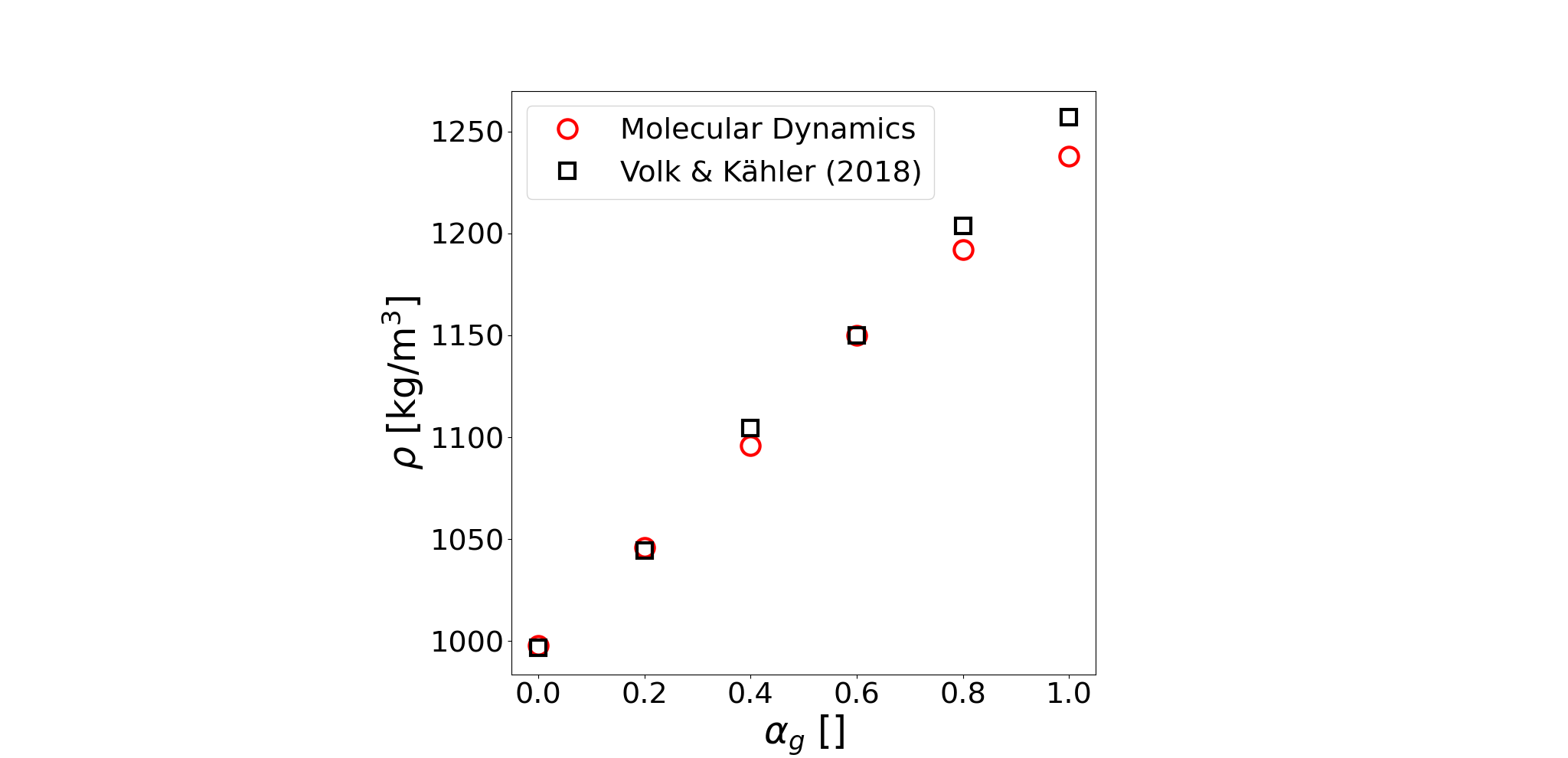}
    \caption{Comparison between density of aqueous glycerol computed from molecular dynamics and experimental values \cite{volk2018glydens}.}
    \label{fig:density}
\end{figure}

Figure \ref{fig:density} reports the average density as a function of the mass fraction of glycerol and a comparison with the experimental data by Volk and Kähler \cite{volk2018glydens}.

\vspace{\abovedisplayskip}
\begin{minipage}[htbp]{\textwidth}
    \begin{minipage}[b]{0.49\textwidth}
        \centering
        \includegraphics[trim={15cm 0cm 15cm 0cm},clip,width=0.9\textwidth]{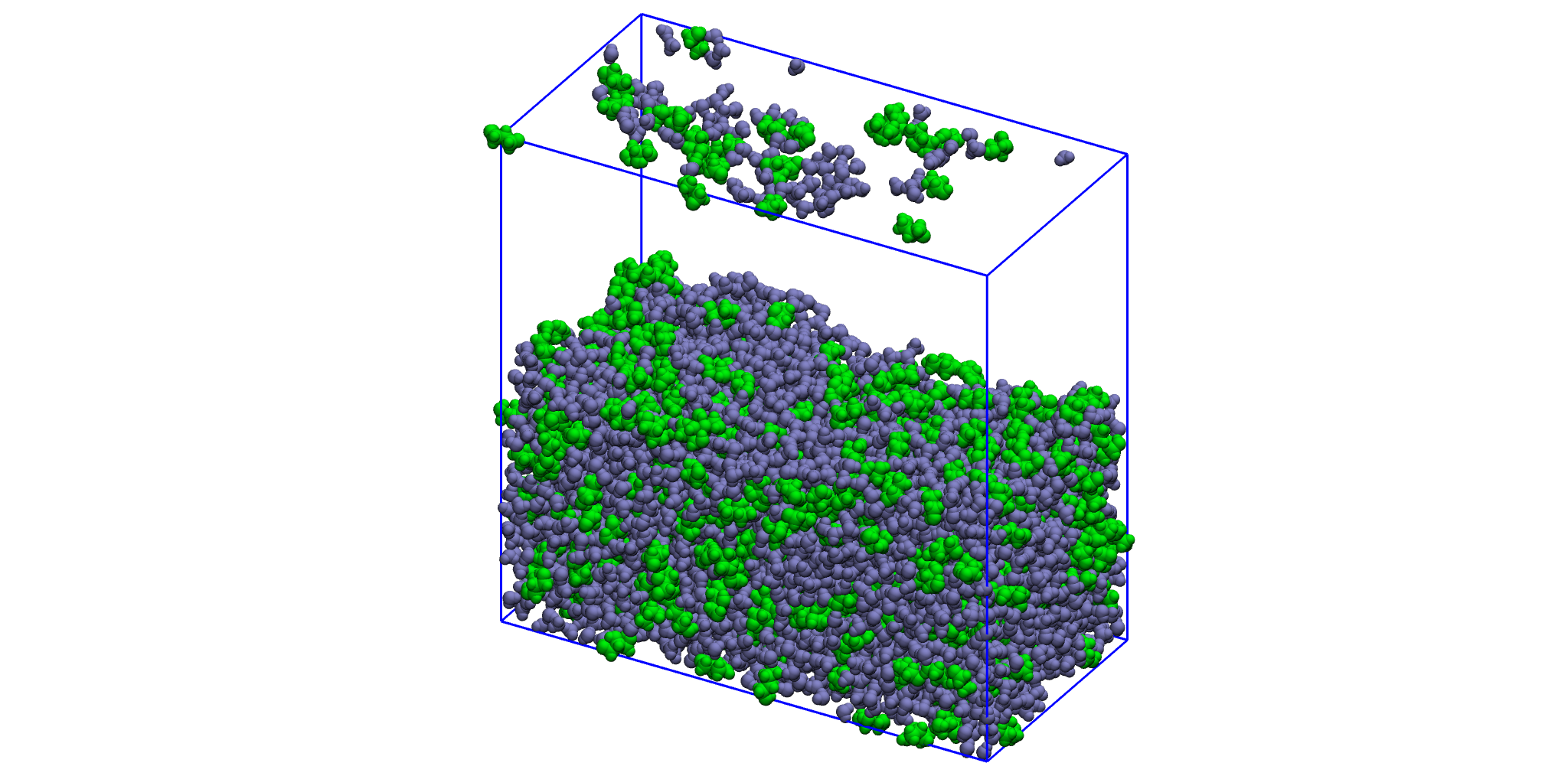}
    \end{minipage}
    \hspace{-0.07\textwidth}
    \begin{minipage}[b]{0.49\textwidth}
        \centering
        \begin{tabular}{ l | l | l }
            $\alpha_g$ & SASA water & SASA glycerol \\
                & [nm$^2$] & [nm$^2$] \\
            \hline
            0.2 & 123.3$\pm$0.6 & 31.3$\pm$0.5 \\  
            0.4 & 90.9$\pm$0.6 & 54.0$\pm$0.4 \\  
            0.6 & 61.6$\pm$0.6 & 88.1$\pm$0.6 \\  
            0.8 & 30.1$\pm$0.3 & 116.4$\pm$0.3
        \end{tabular}
        \vspace{0.5cm}
    \end{minipage}
    \captionof{figure}{Left: visualization of a system of type SLAB. Right: solvent accessible surface area of water and glycerol w.r.t. the interface with vapor.}
    \label{fig:param2}
\end{minipage}
\vspace{\belowdisplayskip}

The SASA (solvent accessible surface area) is computed from systems of type SLAB using the command \texttt{gmx sasa}; we refer to Gromacs documentation for a detailed explanation of the command functionality \cite{gmxdoc}. 

\begin{table}
    \centering
    \begin{tabular}{ r | r || l l l l | c | c }
    & & $r_c=$ & & & & PME & Experimental \\
    & $\alpha_g$ & 1.0 nm & 1.2 nm & 1.4 nm & 1.6 nm & Lennard-Jones & from \cite{takamura2012glycerol} \\
    \hline
    $\sigma$ [10$^{-2}$ Pa$\cdot$m] & 0   & 5.59$\pm$0.03 &5.80$\pm$0.01 & 5.89$\pm$0.01 & 5.97$\pm$0.01 & 6.22$\pm$0.06 & 7.22 \\
    & 0.2	& 5.56$\pm$0.08 & 5.79$\pm$0.01 & 5.91$\pm$0.02 & 6.03$\pm$0.02 & 6.10$\pm$0.03 & 7.1 \\
    & 0.4	& 5.56$\pm$0.04 & 5.85$\pm$0.03	& 6.03$\pm$0.03 & 6.16$\pm$0.02	& 6.50$\pm$0.02 & 6.91 \\
    & 0.6	& 5.60$\pm$0.12 & 5.90$\pm$0.02	& 6.05$\pm$0.04 & 6.20$\pm$0.05	& 6.61$\pm$0.06 & 6.77 \\
    & 0.8	& 5.49$\pm$0.24 & 5.89$\pm$0.04	& 6.10$\pm$0.04 & 6.20$\pm$0.06	& 6.57$\pm$0.06 & 6.67 \\
    & 1	& 5.89$\pm$0.29 & 5.62$\pm$0.20	&6.10$\pm$0.21 & 6.24$\pm$0.17	& 6.53$\pm$0.29 & 6.3 \\
    \end{tabular}
    \caption{Liquid-vapour surface tension for different glycerol concentrations and different Lennard-Jones cutoff radii, compared to PME Lennard-Jones and experimental results.}
    \label{tab:surftens-cutoff}
\end{table}

\begin{figure}[htbp]
    \centering
    \includegraphics[width=0.50\textwidth,trim={0 0 0 0},clip]{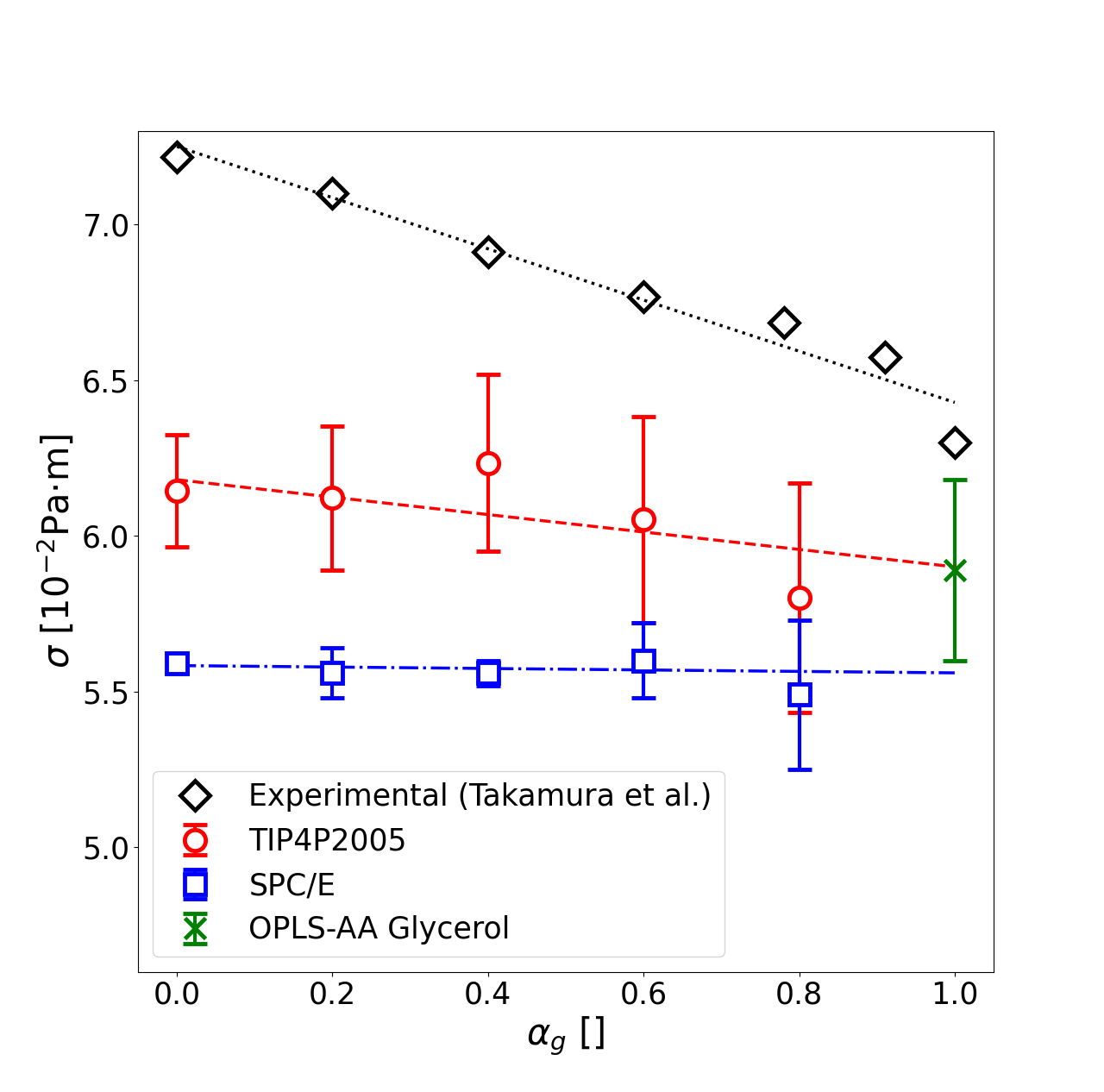}
    \caption{Surface tension of aqueous glycerol solutions for SPC/E and TIP4P/2005 with OPLS-AA glycerol, compared to experimental results from \cite{takamura2012glycerol}.}
    \label{fig:surftens-tip4-exp}
\end{figure}

The calculation of surface tension may be inaccurate due to the long-range treatment of dispersion forces or due to limitations in the water model. Table \ref{tab:surftens-cutoff} reports the effect of the real-space truncation of the Lennard-Jones potential, in relation to the full PME calculation and the experimental results. Figure \ref{fig:surftens-tip4-exp} compares the results for SPC/E with the ones obtained with TIP4P/2005, again alongside experimental data \cite{takamura2012glycerol}.

\begin{figure}[htbp]
    \centering
    \includegraphics[width=0.48\textwidth,trim={0 0 0 0},clip]{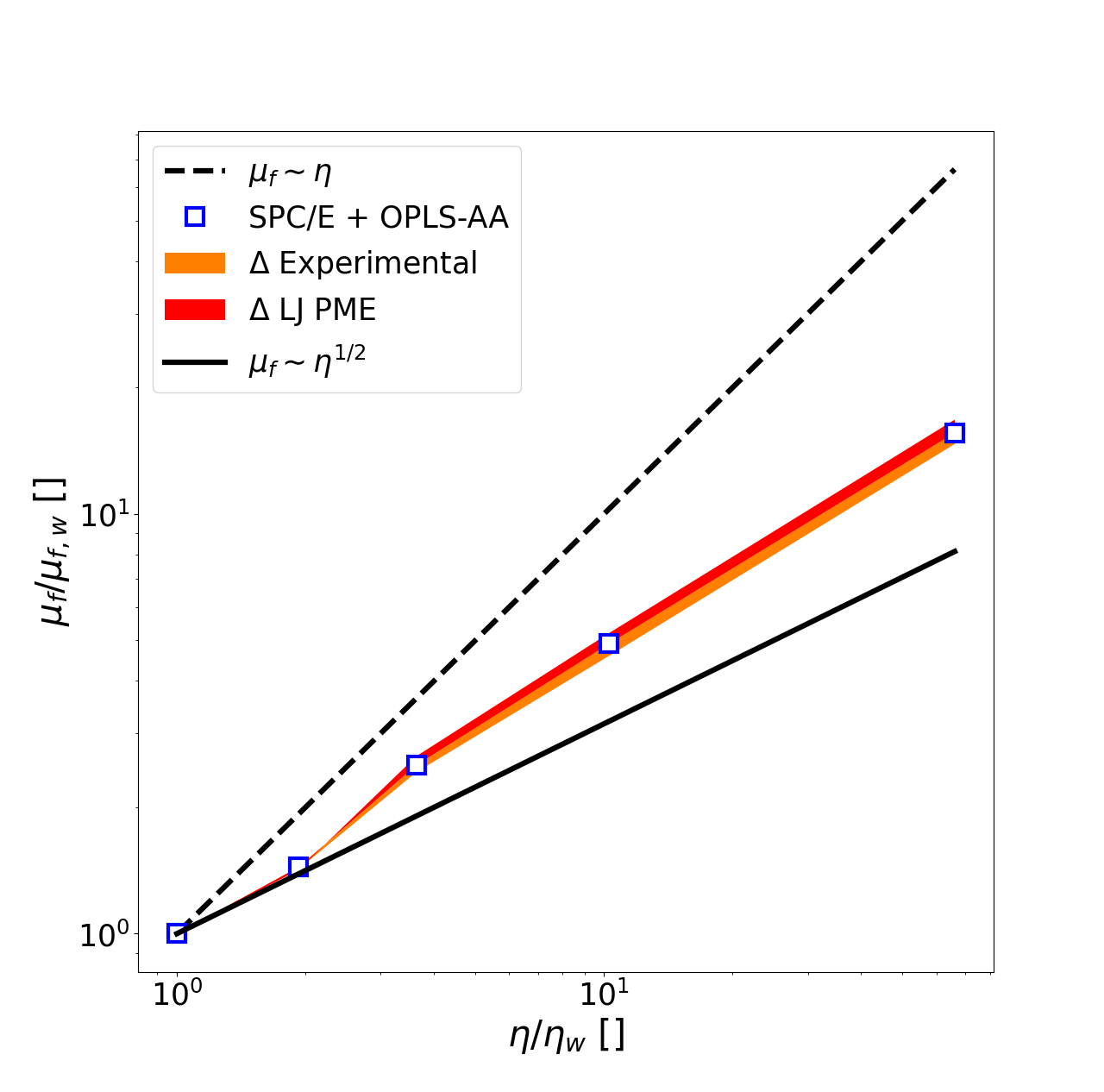}
    \includegraphics[width=0.48\textwidth,trim={0 0 0 0},clip]{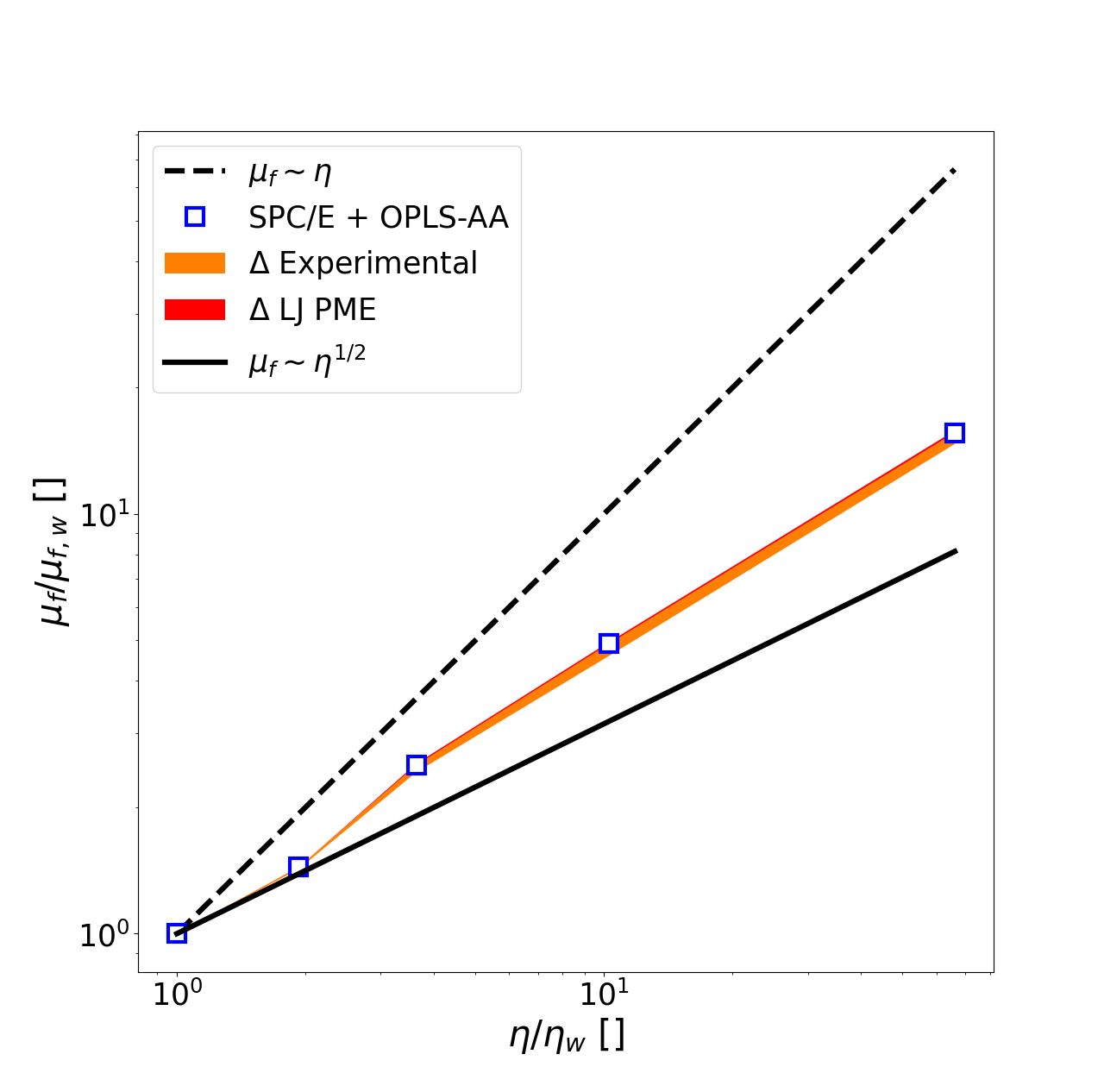}
    \caption{Sensitivity of the scaling between contact line friction and viscosity due to the effect of real-space Lennard-Jones cutoff on surface tension. Left: case (i), right:case (ii); see equation \ref{eq:sensitivity-cl-friction}.}
    \label{fig:pme-lj}
\end{figure}

\begin{figure}[htbp]
    \centering
    \includegraphics[width=0.48\textwidth,trim={0 0 0 0},clip]{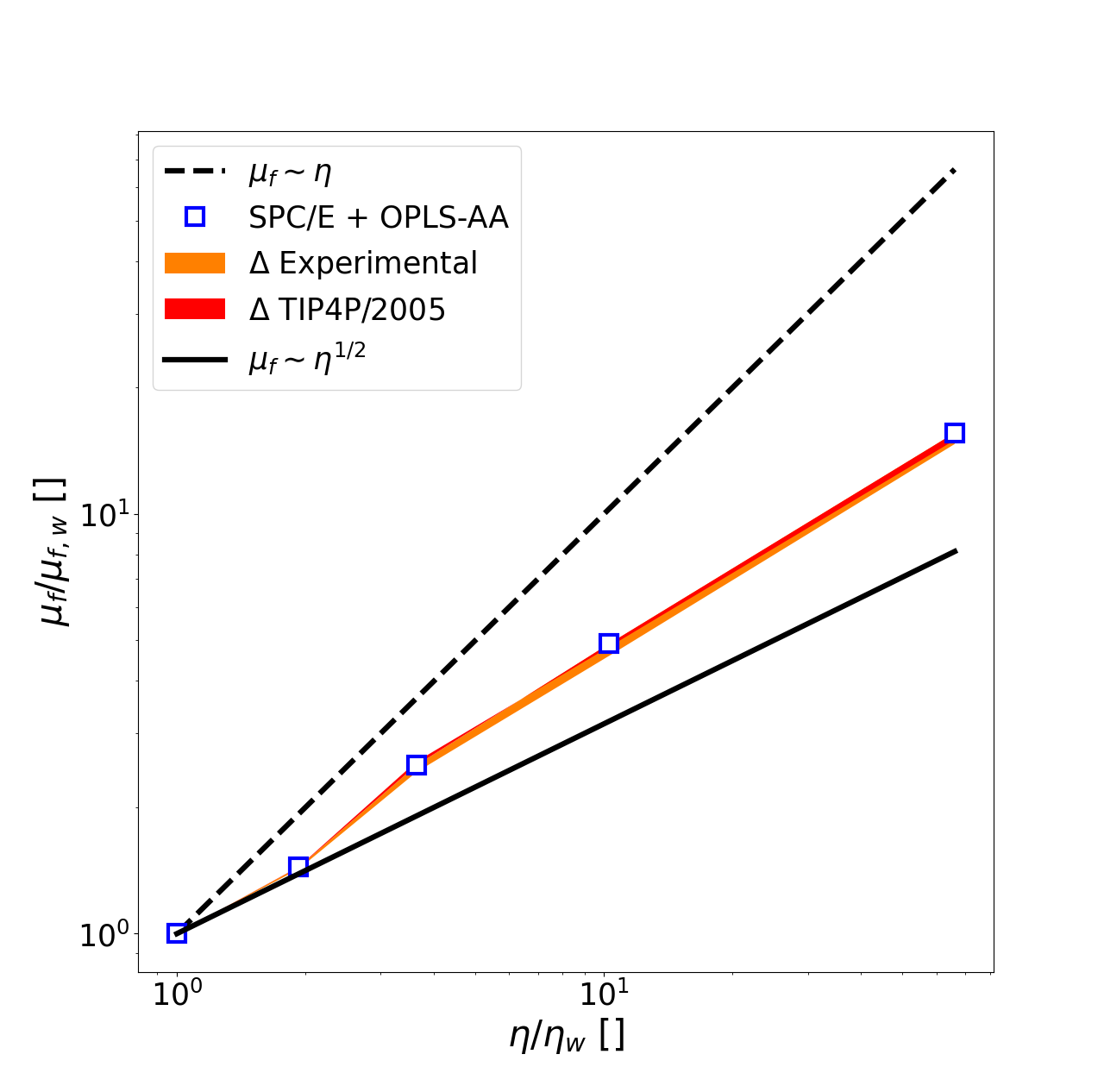}
    \includegraphics[width=0.48\textwidth,trim={0 0 0 0},clip]{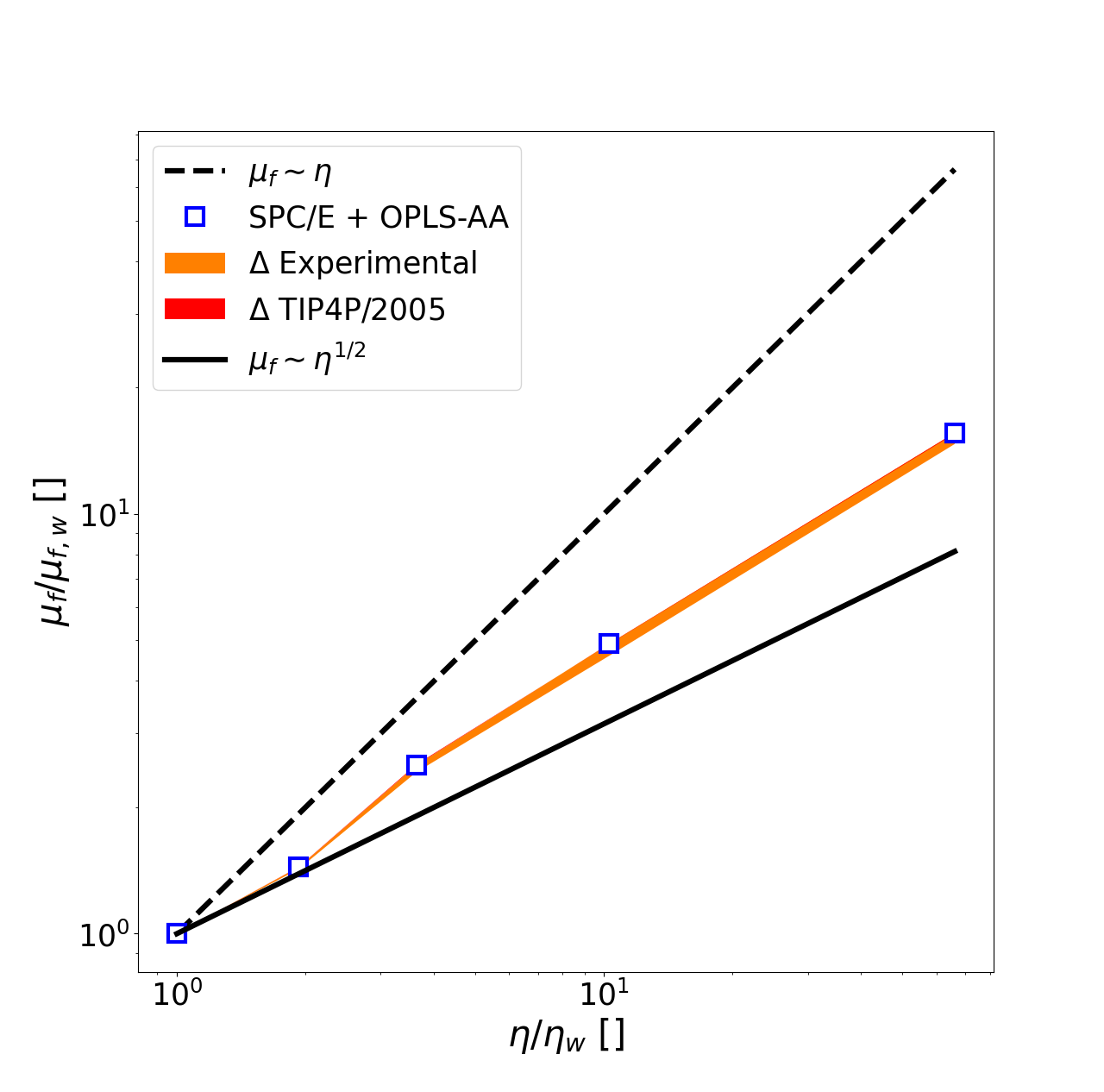}
    \caption{Sensitivity of the scaling between contact line friction and viscosity due to the choice of water model. Left: case (i), right:case (ii); see equation \ref{eq:sensitivity-cl-friction}.}
    \label{fig:sensitivity}
\end{figure}

We have quantified how sensitive the estimate of contact line friction (from equation 1 in the main article) is to the change of surface tension, assuming the spreading dynamics remains the same and fixing either (i) the equilibrium contact angle, or (ii) the difference between solid-vapour and solid-liquid surface energies ($\sigma_{SV}-\sigma_{SL}$). We then estimated the error in the calculation of contact line friction using:
\begin{equation}    \label{eq:sensitivity-cl-friction}
    \begin{dcases}
        \widehat{\mu}_f = \frac{\widehat{\sigma}}{\sigma} \mu_f & \mbox{case (i)} \\
        \widehat{\mu}_f = \mu_f - \expval{\frac{\cos\theta(t)}{u_{cl}(t)}}\big(\widehat{\sigma}-\sigma\big) & \mbox{case (ii)} \\
    \end{dcases} \; ,
\end{equation}
where $\widehat{\sigma}$ is either the experimental surface tension or (a) the one computed using TIP4P/2005 and 1 nm LJ cutoff; (b) the one computed using SPC/E and LJ PME . The symbol $\expval{\cdot}$ indicates the average over the spreading process. Figures \ref{fig:sensitivity} and \ref{fig:pme-lj} show the effect of the perturbation of surface tension on the scaling between contact line friction and viscosity (figure 3b in the manuscript under revision), which is negligible. This result is not surprising, since the change of surface tension for different values of $\alpha_g$, computed either from MD or obtained experimentally, is much smaller than the change in viscosity.\\





\section{Best fit of the linear friction model}   \label{app:best-fit}

In this section we illustrate some of the technical aspects of the data processing for systems of type DROP II. The contact line displacement (obtained according to the procedure mentioned in appendix B in the main article) and the spreading time are first made non-dimensional by dividing by the initial radius $R_0=20$ nm and the viscous-capillary time $\tau=303.5$ ps. A characteristic spreading velocity is obtained as $V_0=R_0/\tau$ and it is used to re-obtain dimensional spreading velocities after data processing.
Since the spreading process slows down in time, a uniform sampling of contact line displacements and contact angles produces a distribution of observations biased towards smaller values. In order to alleviate undesirable over-fitting of the long-time dynamics and under-fitting of the fast dynamics, we re-sample observations in order to obtain a uniform distribution of contact line velocities.

\begin{table}
\begin{center}
\begin{tabular}{ l | l || l | l | l | l | l } 
                  & $\alpha_g$ & $\zeta=0.001$ & $\zeta=0.01$ & $\zeta=0.1$ & $\zeta=1$ & $\zeta=10$    \\
\hline 
 $\mu_f/\eta$ []  & 0   & 5.46$\pm$0.07 & 5.45$\pm$0.08 & 5.46$\pm$0.06 & 5.45$\pm$0.03 & 5.45$\pm$0.06 \\ 
                  & 0.2 & 4.08$\pm$0.05 & 4.07$\pm$0.06 & 4.09$\pm$0.06 & 4.17$\pm$0.07 & 4.44$\pm$0.10 \\ 
                  & 0.4 & 3.77$\pm$0.12 & 3.77$\pm$0.10 & 3.78$\pm$0.08 & 3.93$\pm$0.09 & 4.22$\pm$0.07 \\ 
                  & 0.6 & 2.60$\pm$0.07 & 2.61$\pm$0.07 & 2.61$\pm$0.06 & 2.71$\pm$0.06 & 2.82$\pm$0.02 \\ 
                  & 0.8 & 1.28$\pm$0.03 & 1.28$\pm$0.03 & 1.29$\pm$0.04 & 1.34$\pm$0.03 & 1.44$\pm$0.01 \\
\hline
 $\theta_0$ [deg] & 0   & 48.9$\pm$0.4  & 49.0$\pm$0.4  & 48.9$\pm$0.3  & 48.9$\pm$0.1  & 48.9$\pm$0.1 \\ 
                  & 0.2 & 50.2$\pm$0.5  & 50.3$\pm$0.5  & 50.1$\pm$0.5  & 49.6$\pm$0.5  & 47.6$\pm$0.1 \\ 
                  & 0.4 & 50.0$\pm$0.8  & 49.9$\pm$0.6  & 49.8$\pm$0.6  & 49.0$\pm$0.6  & 47.1$\pm$0.3 \\ 
                  & 0.6 & 53.7$\pm$0.7  & 53.6$\pm$0.7  & 53.7$\pm$0.6  & 52.5$\pm$0.6  & 51.4$\pm$0.1 \\ 
                  & 0.8 & 57.3$\pm$0.8  & 57.3$\pm$0.7  & 57.2$\pm$0.9  & 55.8$\pm$0.8  & 53.4$\pm$0.1 \\ 
\end{tabular}
\caption{Results of the uncertainty quantification procedure (average $\pm$ standard deviation) for each of the model parameters, for different weight values.}
\label{tab:uncert}
\end{center}
\end{table}

The contact line friction is obtained by minimizing the loss function presented in equation 9 in the main article. Table \ref{tab:uncert} reports the value of contact line friction and equilibrium contact angle obtained upon changing the value of the penalization coefficient $\zeta$. The error bars refer to the standard deviation obtained by performing bootstrapping with 20\% of the data, repeated 25 times.

\section{Viscosity}     \label{sec:viscosity}

In this last section we provide additional technical details on the calculation of viscosity using Einstein's relation. In order to obtain an accurate estimate of viscosity we follow the approach of Zhang et al. \cite{zhang2015viscosity} and combine averaging over time to averaging over independent realization of the same observable. The integral in time of the off-diagonal stress tensor components used in Einstein's relation behaves as a Browinan process with correlated noise. Computing viscosity from a single realization would require a surprisingly long trajectory. Running several short replicas in parallel is much more time efficient.

Several independent replicas can be sampled from a sufficiently long NVT equilibrium simulation by sequentially running:\\
\texttt{gmx trjconv -dump <time-to-dump> -f traj.trr -s topol.tpr}\\
$\drsh$ \texttt{-o frame-<frame-number>.gro}\\
The integral of the off-diagonal components of the stress tensor can be obtained given the energy file (\texttt{.edr}) by running:\\
\texttt{gmx energy -f ener.edr -vis -evisco -eviscoi}

For an accurate computation we set \texttt{nstcalcenergy=1} in the molecular dynamics configuration file (\texttt{.mdp}), i.e. energy observables are computed every time step. This is necessary in order to correctly sample the velocity autocorrelation function in time. Since we are interested in the long-time behavior of the integral it is unnecessary to output energy every step and it is sufficient to set \texttt{nstenergy=100}.

\begin{figure}[htbp]
    \centering
    \includegraphics[width=1\textwidth,trim={0 0 0 0}, clip]{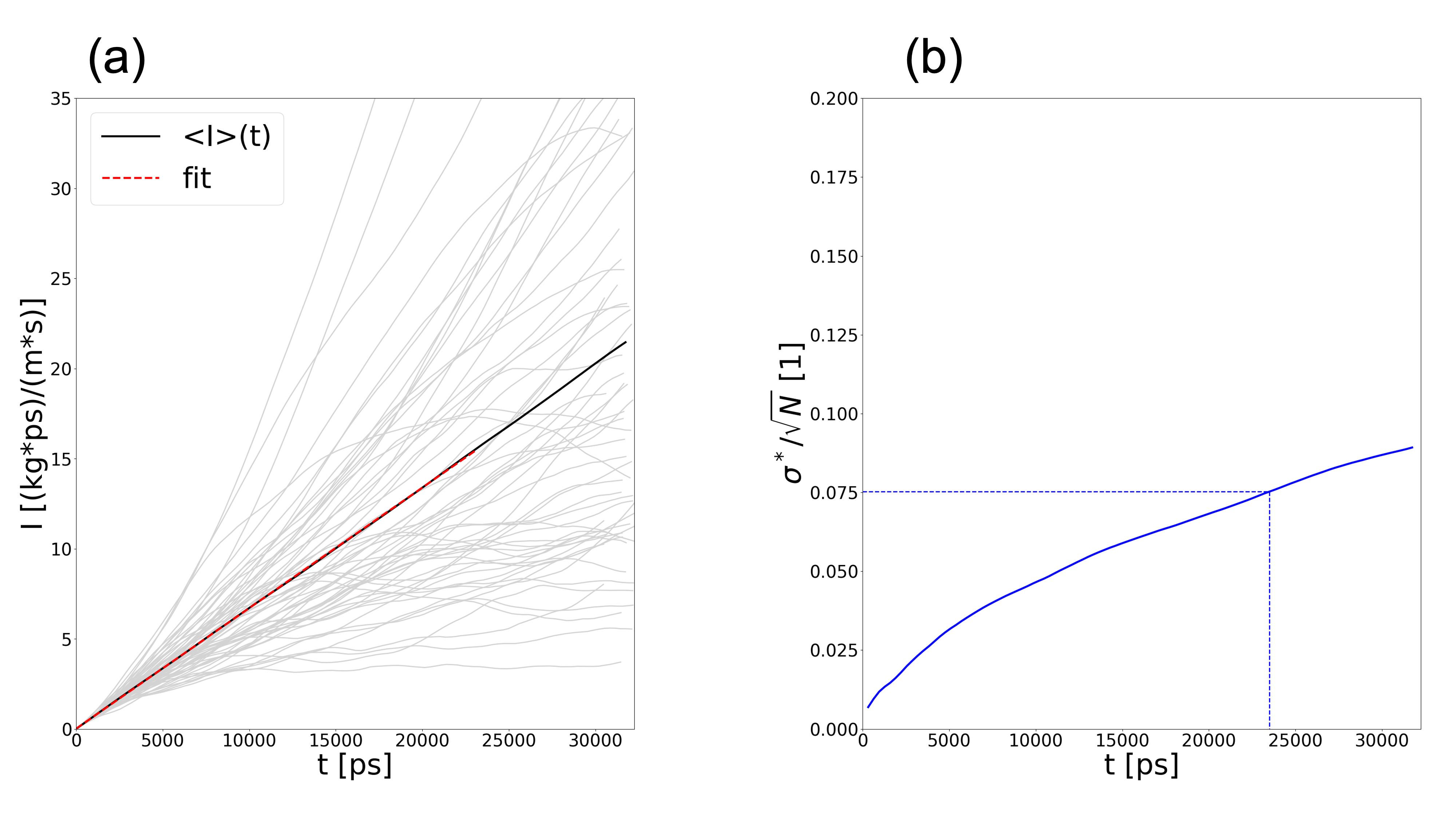}
    \caption{(a) Integrals of the off-diagonal components of the stress tensor for each replica (light gray), their average (solid black line) and its linear fit (dashed red line); results are for for $\alpha_g=0$. (b): Estimate of the relative standard error computed among different realizations, for the same glycerol mass fraction; the error threshold is indicated with a blue dashed line.}
    \label{fig:visco1}
\end{figure}

\begin{figure}[htbp]
    \centering
    \includegraphics[width=1\textwidth,trim={0 0 0 0}, clip]{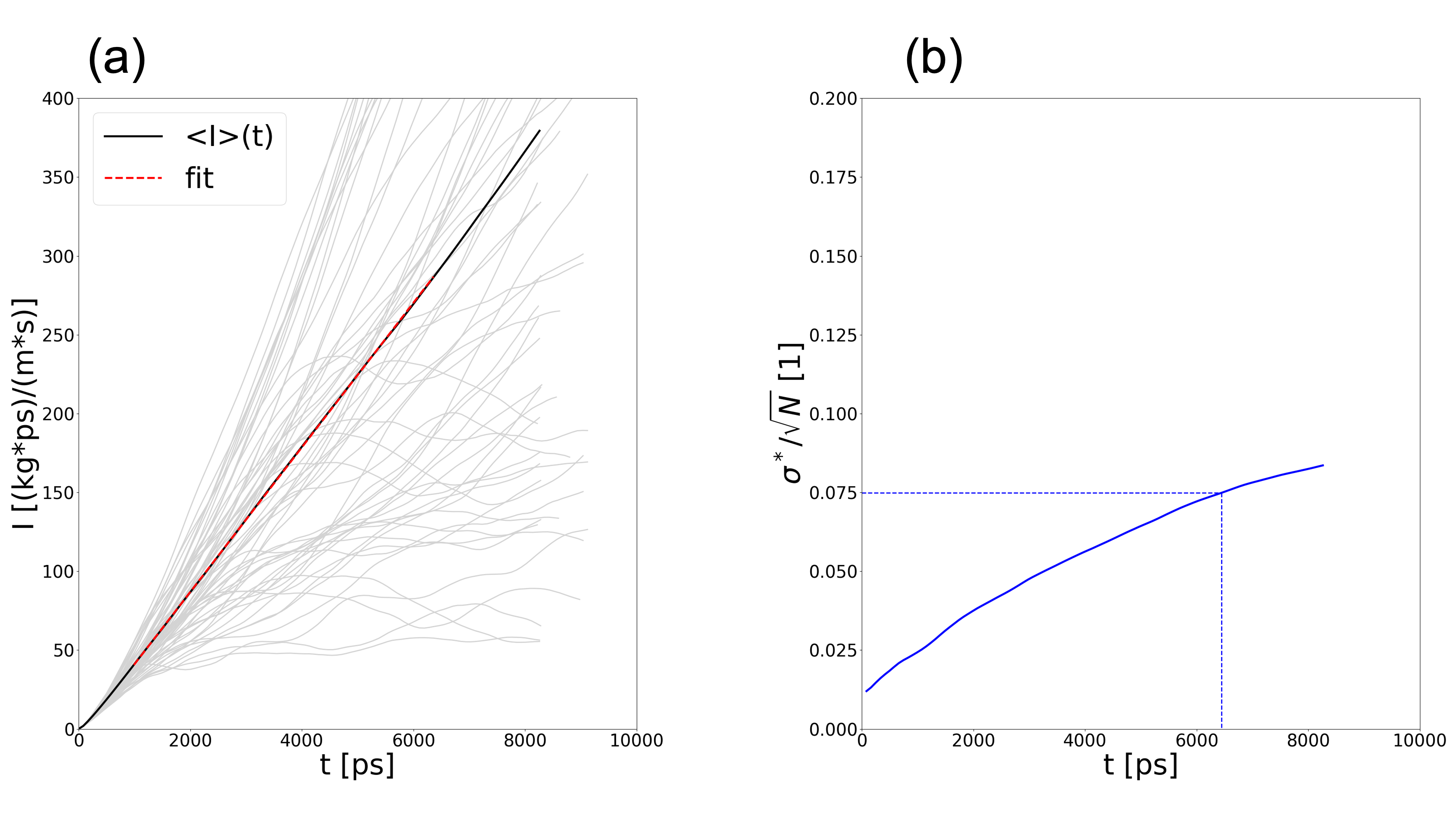}
    \caption{Same quantities as figure \ref{fig:visco1}, but for $\alpha_g=0.8$.}
    \label{fig:visco2}
\end{figure}

Figures \ref{fig:visco1}a and \ref{fig:visco2}a report the integral returned by \texttt{gmx energy -eviscoi} for each replica and its average, for $\alpha_g=0$ and $\alpha_g=0.8$. It can be noted how the Brownian behavior of the observable leads to more uncertain estimates for longer times. We compute the relative standard error for each fixed time frame across 60 replicas and impose an arbitrary threshold of $7.5\%$. 

\begin{figure}[htbp]
    \centering
    \includegraphics[width=1\textwidth,trim={0 0 0 0}, clip]{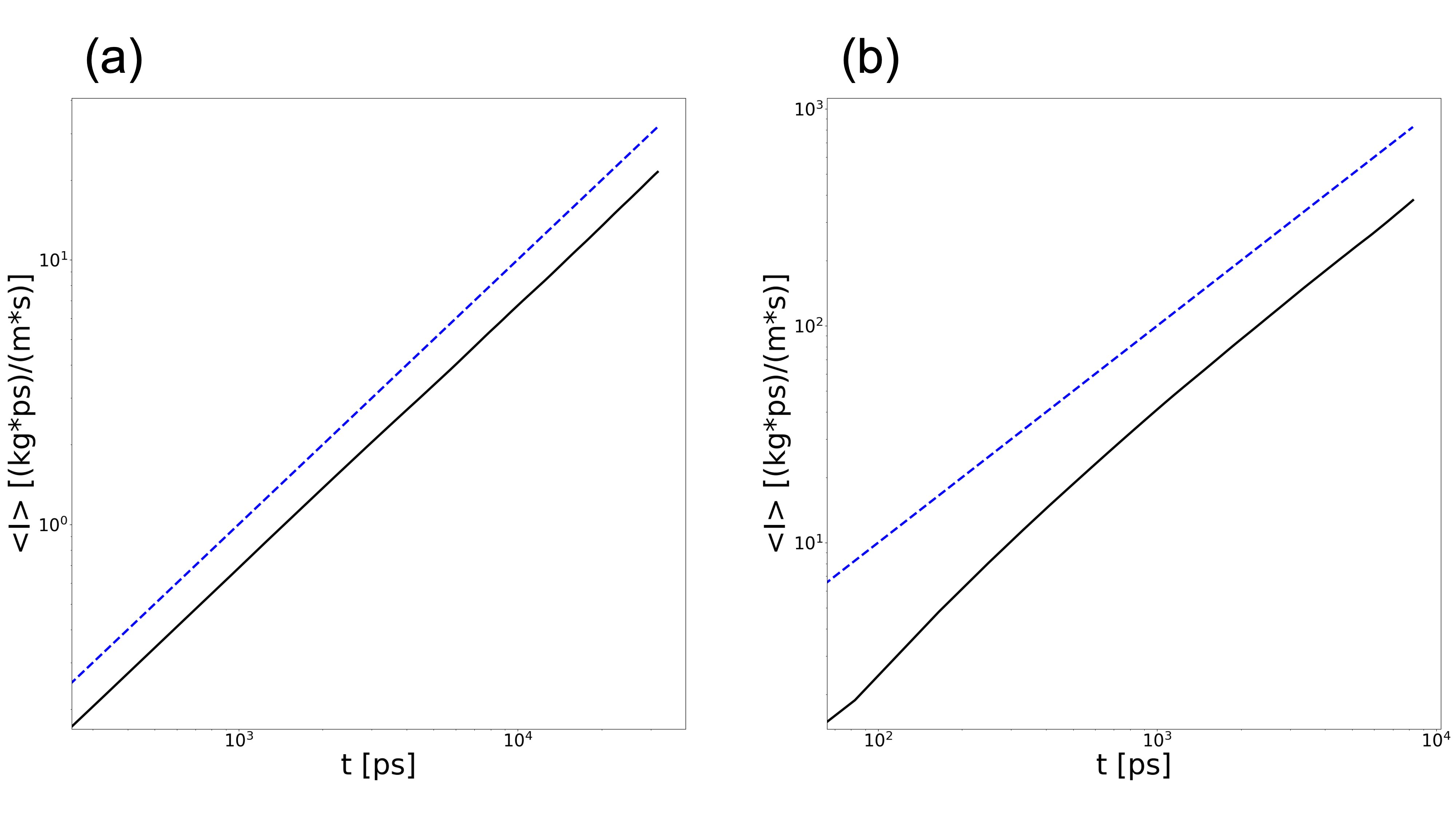}
    \caption{Logarithmic plots of the average integral over time (solid black lines), compared to the expected long-time linear trend (dashed blue line). (a): $\alpha_g=0$, (b): $\alpha_g=0.8$.}
    \label{fig:visco3}
\end{figure}

Viscosity is obtained via linear regression of the integral up to the time when the estimated relative standard error reaches the threshold. The regression is inversely weighted on the relative standard error. We also discard an initial time window corresponding to a `spin-up time': it is know from linear response theory that the long-time behavior in time of the average of the integral should be linear; the initial super-linear growth rate regime is identified by visual inspection (figure \ref{fig:visco3}) and discarded. The uncertainty on the value of the shear viscosity coefficient is estimated by randomly removing 20\% of replicas and repeating the coefficient estimation 20 times.

\bibliography{apssamp}